\documentclass[preprint,11pt]{elsarticle}
\usepackage[english]{babel}
\usepackage[latin1]{inputenc}
\usepackage[T1]{fontenc}
\usepackage{graphicx,color}
\usepackage{amsmath}
\usepackage{amssymb}
\usepackage{array}
\usepackage[margin=1.8cm]{geometry}
\usepackage{listings}
\begin{document}

\begin{frontmatter}

\title{\Large\textsc{An efficient and portable SIMD algorithm for charge/current deposition in Particle-In-Cell codes}}
\author[label1,label2]{H. Vincenti\corref{cor1}}
\ead{hvincenti@lbl.gov}

\author[label1]{R. Lehe}
\ead{rlehe@lbl.gov}

\author[label3]{R. Sasanka}
\ead{ruchira.sasanka@intel.com}

\author[label1]{J-L. Vay}
\ead{jlvay@lbl.gov}

\address[label1]{Lawrence Berkeley National Laboratory, 1 cyclotron road, Berkeley, California, USA}
\address[label2]{Lasers Interactions and Dynamics Laboratory (LIDyL), Commissariat à l'Energie Atomique, Gif-Sur-Yvette, France}
\address[label3]{Intel corporation,  Oregon,  USA}

\cortext[cor1]{Corresponding author}

\begin{abstract}
In current computer architectures, data movement (from die to network) is by far the most energy consuming part of an algorithm (10pJ/word on-die to 10,000pJ/word on the network). To increase memory locality at the hardware level and reduce energy consumption related to data movement, future exascale computers tend to use more and more cores on each compute nodes ("fat nodes") that will have a reduced clock speed to allow for efficient cooling. To compensate for frequency decrease, machine vendors are making use of long SIMD instruction registers that are able to process multiple data with one arithmetic operator in one clock cycle. SIMD register length is expected to double every four years. As a consequence, Particle-In-Cell (PIC) codes will have to achieve good vectorization to fully take advantage of these upcoming architectures. In this paper, we present a new algorithm that allows for efficient and portable SIMD vectorization of current/charge deposition routines that are, along with the field gathering routines, among the most time consuming parts of the PIC algorithm. Our new algorithm uses a particular data structure that takes into account memory alignement constraints and avoids gather/scatter instructions that can significantly affect vectorization performances on current CPUs. The new algorithm was successfully implemented in the 3D skeleton PIC code PICSAR and tested on Haswell Xeon processors (AVX2-256 bits wide data registers). Results show a factor of $\times 2$ to $\times 2.5$ speed-up in double precision for particle shape factor of order $1$ to $3$. The new algorithm can be applied as is on future KNL (Knights Landing) architectures that will include AVX-512 instruction sets with 512 bits register lengths (8 doubles/16 singles). 
\end{abstract}

\end{frontmatter}

\newpage

\section{Introduction} 






\subsection{Challenges for porting PIC codes on exascale architectures: importance of vectorization}

Achieving exascale computing facilities in the next decade will be a great challenge in terms of energy consumption and will imply hardware and software developments that directly impact our way of implementing PIC codes \cite{Birdsall}. 

\begin{table}[h]
\centering
\begin{tabular}{|c|c|c|}
\hline
Operation & Energy cost & Year\\
\hline
DP FMADD flop & $11pJ$ &2019\\
\hline
Cross-die per word access & $24pJ$ &2019\\
\hline
DP DRAM read to register & $4800 pJ$ &2015\\
\hline
DP word transmit to neighbour & $7500 pJ$ & 2015\\
\hline
DP word transmit across system & $9000 pJ$& 2015\\
\hline
\end{tabular}
\caption{Energy consumption of different operations taken from \cite{DARPA2008}. The die hereby refers to the integrated circuit board made of semi-conductor materials that usually holds the functional units and fast memories (first levels of cache). This table shows the energy required to achieve different operations on current (Year 2015) and future  (Year 2019) computer architectures. DP stands for Double Precision, FMADD for Fused Multiply ADD and DRAM for Dynamic Random Access Memory.  }
\label{Table1}
\end{table}

Table \ref{Table1} shows the energy required to perform different operations ranging from arithmetic operations (fused multiply add or FMADD) to on-die memory/DRAM/Socket/Network memory accesses. As $1$pJ/flop/s is equivalent to $1$MW for exascale machines delivering $1$ exaflop ($10^{18}$ flops/sec), this simple table shows that as we go off the die, the cost of memory accesses and data movement becomes prohibitive and much more important than simple arithmetic operations.  In addition to this energy limitation, the draconian reduction in power/flop and per byte will make data movement less reliable and more sensitive to noise, which also push towards an increase in data locality in our applications. 

At the hardware level, part of this problem of memory locality was progressively adressed in the past few years by limiting costly network communications and grouping more computing ressources that share the same memory ("fat nodes"). However, partly due to cooling issues, grouping more and more of these computing units will imply a reduction of their clock speed. To compensate for the reduction of computing power due to clock speed, future CPUs will have much wider data registers that can process or "vectorize" multiple data in a single clock cycle (Single Instruction Multiple Data or SIMD). 

At the software level, programmers will need to modify algorithms so that they achieve both memory locality and efficient vectorization to fully exploit the potential of future exascale computing architectures.

\subsection{Need for portable vectorized routines}

In a standard PIC code, the most time consuming routines are current/charge deposition from particles to the grid and field gathering from the grid to particles. These two operations usually account for more than $80\%$ of the execution time. 
Several portable deposition algorithms were developed and successfully implemented on past generations' vector machines (e.g. CRAY, NEC) \cite{NOY1985,horowitz1987,sh1987,heron1989,paruolo1990}. However, these algorithms do not give good performance on current SIMD architectures, that have new constraints in terms of memory alignement and data layout in memory. 

To the authors' knowledge, most of the vector deposition routines proposed in contemporary PIC codes use compiler based directives or even C++ Intel intrinsics in the particular case of the Intel compiler,  to increase vectorization efficiency (e.g. \cite{Fonseca2013}). However, these solutions are not portable and require code re-writing for each new architecture. 

\subsection{Paper outline}

In this paper, we propose a portable algorithm for the direct deposition of current or charge from macro particles onto a grid, which gives good performances on SIMD machines. The paper is divided into four parts: 
\begin{enumerate}[(i)]
\item in section $2$, we quickly introduce the standalone 3D skeleton electromagnetic PIC code PICSAR-EM3D in which we implemented the different vector versions of the deposition routines presented in this paper,
\item in section $3$, we quickly remind the scalar deposition routine and show why it cannot be vectorized as is by the compiler. Then, we introduce a vector algorithm that performed well on former Cray vector machines but give poor performances on current SIMD machines. By carefully analyzing the bottlenecks of the old vector routine on current SIMD machines, we will derive a new vector routine that gives much better performances,
\item in section $4$ we present the new vector routines that was developed, based on the analysis in section $3$, 
\item in section $5$, the new vector routines are benchmarked on the new Cori machine at the U.S. National Energy Research Supercomputer Center (NERSC) \cite{Nersc}. 
\end{enumerate}

\section{The PICSAR-EM3D PIC kernel}

PICSAR-EM3D is a standalone "skeleton" PIC kernel written in Fortran 90 that was built using the main electromagnetic PIC routines (current deposition, particle pusher, field gathering, Yee field solver) of the framework WARP \cite{Warp}. As WARP is a complex mix of Fortran 90, C and Python, PICSAR-EM3D provides an essential testbed for exploring PIC codes algorithms with multi-level parallelism for emerging and future exascale architectures. All the high performance carpentry and data structures in the code have been redesigned for best performance on emerging architectures, and tested on NERSC supercomputers (CRAY XC30  Edison and testbed with Intel Knight's Corner coprocessors Babbage). 

\subsection{PIC algorithm}
PICSAR-EM3D contains the essential features of the standard explicit electromagnetic PIC main loop:  

\begin{enumerate}[(i)]
\item Maxwell solver using arbitrary order finite-difference scheme (staggered or centered), 
\item Field gathering routines including high-order particle shape factors (order 1 - CIC, order 2 - TSC and order 3 - QSP),
\item Boris particle pusher, 
\item Most common types of current depositions : Morse-Nielson  deposition \cite{Birdsall} (also known as direct $\rho \textbf{v}$ current deposition) and  Esirkepov \cite{Esirkepov2001} (charge conserving) schemes. The current and charge deposition routines support high-order particle shape factors (1 to 3). 
\end{enumerate} 

\subsection{High performance features}

Many high performance features have already been included in PICSAR-EM3D. In the following, we give a quick overview of the main improvements that brought significant speed-up of the code and that are of interest for the remainder of this paper. A more comprehensive description  of the code and its performances will be presented in another paper. 

\subsubsection{Particle tiling for memory locality}

Field gathering (interpolation of field values from the grid to particle positions) and current/charge deposition (deposition of particle quantities to adjacent grid nodes) account for more than 80\% of the total execution time of the code. In the deposition routines for instance, the code loops over all particles and deposit their charges/currents on the grid. 

\begin{figure}[h!]
\centering
\includegraphics[width=0.6\linewidth]{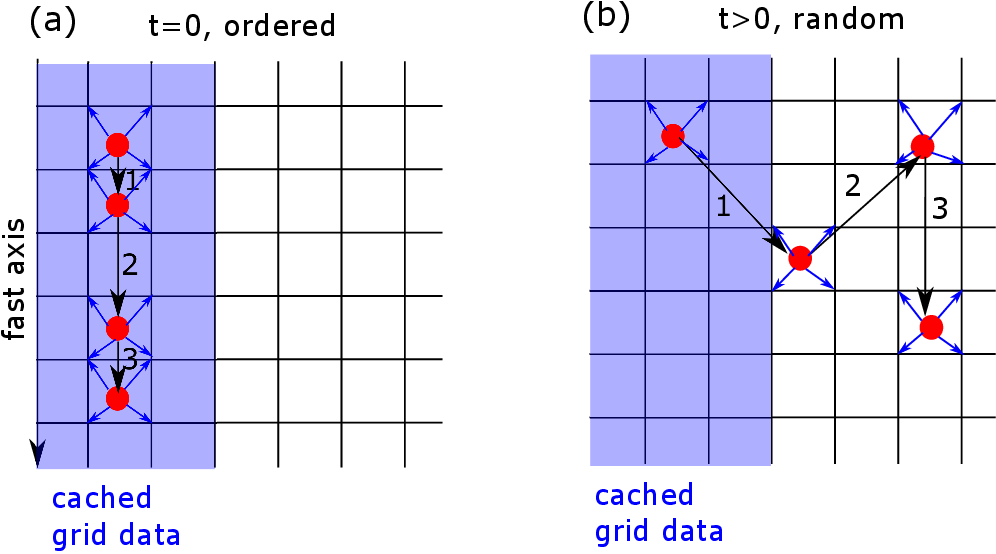}
\caption{Importance of cache reuse in deposition routines. Illustration is given in 2D geometry for clarity, with CIC (linear) particle shapes. Panel (a) shows a typical layout at initialization (t=0) where particles are ordered along the "fast" axis of the grid, corresponding to grid cells (blue area) that are contiguous in memory. The loop on particles is illustrated with arrows and index of the loop with numbers 1 to 3.  Using direct deposition, each particle (red point) deposits (blue arrows) its charge/current to the nearest vertices (4 in 2D and 8 in 3D for CIC particle shapes). Panel (b) illustrates the random case (at t>0) where particles are randomly distributed on the grid. As the algorithms loops over particles, it often requires access to uncached grid data, which then results in a substantial number of cache misses. }
\label{CacheReuse}
\end{figure}

One major bottleneck that might arise in these routines and can significantly affect overall performance is cache reuse.

Indeed, at the beginning of the simulations (cf. Fig. \ref{CacheReuse} (a)) particles are typically ordered along the "fast" axis ("sorted case") that corresponds to parts of the grid that are contiguously located in memory.  As the code loops over particles, it will thus access contiguous grid portions in memory from one particle to another and efficiently reuse cache.

However, as time evolves, the particle distribution often becomes increasingly random, leading to numerous cache misses in the deposition/gathering routines (cf. Fig. \ref{CacheReuse} (b)). This results in a considerable decrease in performance. In 2D geometry, one MPI subdomain usually fits in L2 cache (256kB to 512 kB per core) but for 3D problems with MPI subdomains handling 100x100x100 grid points, one MPI subdomain does not fit in cache anymore and random particle distribution of particles can lead to performance bottlenecks. 

\begin{figure}[h!]
\centering
\includegraphics[width=0.6\linewidth]{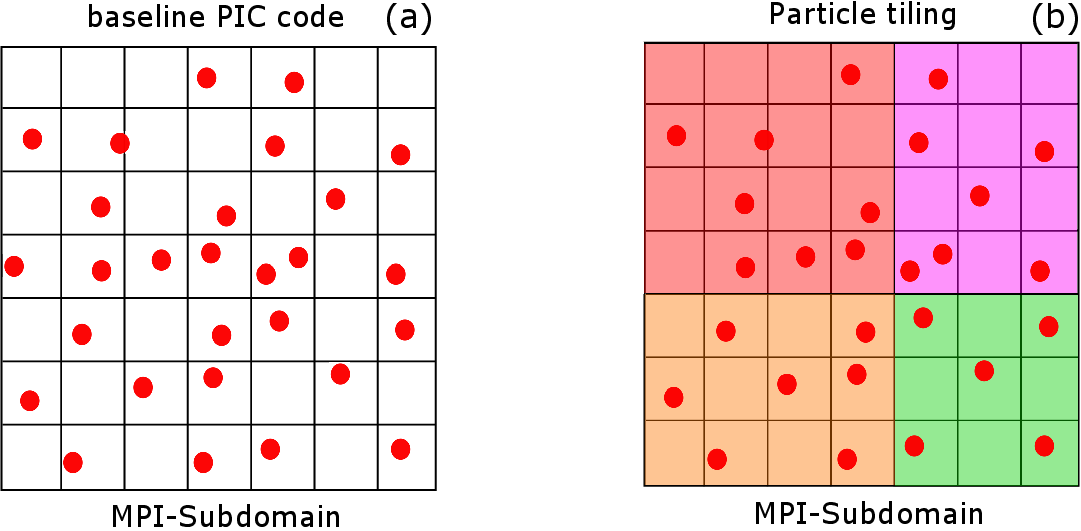}
\label{TilingPrinciple}
\caption{Particle tiling for efficient cache reuse. Panel (a) shows the usual configuration used in standard codes. There is one big array for particles for each MPI subdomain. Panel (b) shows the data structure used in PICSAR. Particles are grouped in tiles that fit in cache, allowing for efficient cache reuse during deposition/gathering routines. }
\end{figure}

 To solve this problem and achieve good memory locality, we implemented particle tiling in PICSAR-EM3D. Particles are placed in tiles that fit in  cache (cf. Fig. \ref{CacheReuse} (b)). In the code, a tile is represented by a structure of array \textit{Type(particle\_tile)} that contains arrays of particle quantities (positions, velocity and weight). All the tiles are represented by a 3D Fortran array \textit{array\_of\_tiles(:,:,:)} of type \textit{particle\_tile} in the code. Our data structure is thus very different from the one in \cite{Decyk2014} which uses one large Fortran $ppart(1:ndims,1:nppmax,1:ntiles)$ array for all particles and tiles, where $ndims$ is the number of particle attributes (e.g positions $x$, $y$, $z$), $nppmax$ the maximum number of particles in a tile and $ntiles$ the number of tiles. There are two reasons behind our choice:  
 
\begin{enumerate}[(i)]
\item if one tile has much more particles than others, we considerably save memory by using our derived type compared to the array $ppart$. Indeed, in the latter case, if one tile has much more particles $np$ than others, we would still need to choose $nppmax\geqslant np$ for all the tiles,
\item any tile can be resized as needed independently, without the need for reallocating the entire array of tiles. 
\end{enumerate}
  
 Performance improvements of the whole code are reported on table \ref{PerformanceTiling} for tests performed on Intel Ivy Bridge (Cray XC30 Edison machine at NERSC). These tests show a speed-up of x3 in case of a random particle distribution. Cache reuse using tiling reaches $99\%$. The optimal  tile size ranges empirically between 8x8x8 cells to 10x10x10 cells. As will be shown later in the paper, having good cache reuse is crucial to increasing the flop/byte ratio of the proposed algorithm and obtaining improvements using vectorization. 
 
 \begin{table}[h!]
 \centering
 \begin{tabular}{|c|c|c|}
 \hline
Tile size& Speed-up & L1 and L2 Cache reuse\\
\hline
$1\times1\times1$& $\times 1$ & $85\%$\\
\hline
$10\times10\times10$& $\times 3$ & $99\%$\\
\hline
 \end{tabular}
 \caption{Speed-up of the whole PIC code brought by particle tiling. Tests were performed using a 100x100x100 grid with 10 particle per cells. Particles are randomly distributed on the grid and have an initial temperature of $10 keV$. The reference time corresponds to the standard case of 1x1x1 tile. The tests were performed on one MPI process and a single socket, on the Edison cluster at NERSC.}
  \label{PerformanceTiling}
 \end{table}
 
Notice that at each time step, the particles of each tile are advanced and then exchanged between tiles. As particles move less than one cell at each time step, the amount of particles exchanged between tiles at each time step is low for typical tiles' sizes. (The surface/volume ratio decreases with tile size.) As a consequence, particle exchanges between tiles account in practice for a very small percentage of the total PIC loop (a few percents). Our particle exchange algorithm differs from the one used in  \cite{Decyk2014} in that it avoids copying data into buffers.  In addition, it can be efficiently parallelized using OpenMP (details are beyond the scope for this paper and will be presented in an upcoming publication). 

\subsubsection{Multi-level parallelization}

PICSAR-EM3D also includes the following high performance implementations: 
\begin{enumerate}[(i)]
\item vectorization of deposition and gathering routines, 
\item OpenMP parallelization for intranode parallelisms. Each OpenMP thread handles one tile. As there are  much more tiles than threads in 3D, load balancing can be easily done using the SCHEDULE clause in openMP with the guided attribute, 
\item MPI parallelization for internode parallelism, 
\item MPI communications are overlapped with computations. For particles,  this is done by treating exchanges of particles with border tiles while performing computations on particles in inner tiles, 
\item MPI-IO for fast parallel outputs. 
\end{enumerate}

In the remainder of this paper, we will focus on the vectorization of direct charge/current deposition routines for their simplicity and widespread use in electromagnetic PIC codes. The Esirkepov-like current deposition is not treated in this paper but the techniques used here are very general and should apply in principle to any kind of current deposition. 

\section{Former CRAY vector algorithms and performance challenges on new architectures}

In the following, we focus on the direct $3D$ charge deposition which can be presented in a more concise way than the full $3D$ current deposition. Vectorization methods presented for charge deposition can easily be transposed to current deposition and $3D$ vector algorithms for current deposition can be found in appendix B.

\subsection{Scalar algorithm}

The scalar algorithm for order 1 charge deposition is detailed in listing \ref{ScalarRoutineCIC}.  For each particle index $ip$, this algorithm (see line $5$): 
\begin{enumerate}[(i)]
\item  finds the indices $(j,k,l)$ of the cell containing the particle (lines $11-13$), 
\item computes the weights of the particle at the 8 nearest vertices $w1$ to $w8$ (line $15$-not shown here), 
\item adds charge contribution to the eight nearest vertices $\{(j,k,l),(j+1,k,l),(j,k+1,l),(j+1,k+1,l),(j,k,l+1),(j+1,k,l+1),(j,k+1,l+1),(j+1,k+1,l+1)\}$ of the current cell  $(j,k,l)$ (see lines $18-25$).  
 \end{enumerate}
\begin{lstlisting}[frame=single,caption=Scalar charge deposition routine for CIC particle shape factors,label=ScalarRoutineCIC,basicstyle=\ttfamily\footnotesize] 
SUBROUTINE depose_rho_scalar_1_1_1(...)
	! Declaration and init
	! ...........7
	! Loop on particles
        DO ip=1,np
            ! --- computes current position in grid units
            x = (xp(ip)-xmin)*dxi
            y = (yp(ip)-ymin)*dyi
            z = (zp(ip)-zmin)*dzi
            ! --- finds node of cell containing  particle
            j=floor(x)
            k=floor(y)
            l=floor(z)
            ! --- computes weigths w1..w8
            ........
            ! --- add charge density contributions
            ! --- to the 8 vertices of current cell
            rho(j,k,l)		=rho(j,k,l)	 + w1
            rho(j+1,k,l)	=rho(j+1,k,l)	 + w2
            rho(j,k+1,l)	=rho(j,k+1,l) 	 + w3
            rho(j+1,k+1,l)	=rho(j+1,k+1,l)	 + w4
            rho(j,k,l+1)	=rho(j,k,l+1)	 + w5
            rho(j+1,k,l+1)	=rho(j+1,k,l+1)	 + w6
            rho(j,k+1,l+1)	=rho(j,k+1,l+1)	 + w7
            rho(j+1,k+1,l+1)	=rho(j+1,k+1,l+1)+ w8
        END DO
END SUBROUTINE depose_rho_scalar_1_1_1
\end{lstlisting}

As two different particles $ip_{1}$ and $ip_2$ can contribute to the charge at the same grid nodes, the loop over particles (line $5$) presents a dependency and is thus not vectorizable as is. 

\subsection{Former vector algorithms and new architecture constraints}

Several vector algorithms have already been derived and tuned on former Cray vector machines \cite{NOY1985,horowitz1987,sh1987,heron1989,paruolo1990,Anderson1995}. However, these techniques are not adapted anymore to current architectures and yield very poor results on SIMD machines that  necessitate to comply with the three following constraints in order to enable vector performances: 
\begin{enumerate}[(i)]
\item \textbf{Good cache reuse}. The flop/byte ratio (i.e. cache reuse) in the main loops of the PIC algorithm must be high in order to observe a speed-up with vectorization. Otherwise, if data has to be moved from memory to caches frequently, the performance gain with vectorization can become obscured by the cost of data movement. As we showed earlier, this is ensured by particle tiling in our code,
\item \textbf{Memory alignement}. Data structures in the code need to be aligned and accessed in a contiguous fashion in order to maximize performances.  Modern computers read from or write to a memory address in word-sized chunks of 8 bytes (for 64 bit systems). Data alignment consists in putting the data at a memory address equal to some multiple of the word size, which increases the system's performance due to the way the CPU handles memory. SSE2, AVX and AVX-512 on x86 CPUs do require the data to be 128-bits, 256-bits and 512-bits aligned respectively, and there can be substantial performance advantages from using aligned data on these architectures. Moreover, compilers can generate more optimal vector code when data is known to be aligned in memory. In practice, the compiler can enforce data alignment at given memory boundaries (128, 256 or 512 bits) using compiler flags/directives.  
\item \textbf{Unit-stride read/write}. If data are accessed contiguously in a do loop (unit-stride), the compiler will generate vector single load/store instructions for the data to be processed. Otherwise, if data are accessed randomly or via indirect indexing, the compiler might generate gather/scatter instructions that almost yield sequential performance or worse. Indeed, in case of a gather/scatter, the processor might have to make several different loads/stores from/to memory instead of one load/store, eventually leading to poor vector performances.  
\end{enumerate}

 In the following, we investigate performances of one of the former vector algorithm for CRAY machines \cite{sh1987} and analyze its bottlenecks on SIMD architectures.  This analysis will show a way to improve the vector algorithm and derive a new one that yields significant speed-up over the scalar version. 

\subsection{Example: the Schwarzmeier and Hewit scheme (SH)}

\subsubsection{SH vector deposition routine}

Listing \ref{VecSHRoutineCIC} details the Schwarzmeier and Hewitt (SH) deposition scheme \cite{sh1987} that was implemented in PICSAR-EM3D and tested on Cori supercomputer at NERSC.  In this scheme, the initial loop on particles is done by blocks of lengths $nblk$ (cf. line $5$) and split in two consecutive nested loops: 
\begin{itemize}
\item A first nested loop (line $7$) that computes, for each particle $nn$ of the current block: 
\begin{enumerate}[(i)]
\item its cell position $ind0$ on the mesh (line $13$), 
\item its contribution $ww(1,nn),...,ww(8,nn)$ to the charge at the $8$ vertices of the cell and 
\item the indices $ll(1,nn),...,ll(8,nn)$ of the $8$ nearest vertices in the 1D density array rho (cf. lines $14-19$). 
\end{enumerate}
Notice that 1D indexing is now used for $rho$ to avoid storing three different indices for each one of the 8 vertices. The Fortran integer array $moff(1:8)$ gives the indices of the 8 vertices with respect to the cell index $ind0$ in the 1D array $rho$. The loop at line $7$ has no dependencies and is vectorized using the portable \textit{\$OMP SIMD} directive. 
\item A second nested loop (line $23$) that adds the contribution of each one of the $nblk$ particles to the $8$ nearest vertices of their cell (line $26$). As one particle adds its contribution to eight different vertices, the loop on the vertices at line $25$ has no dependency and can also be vectorized using the \textit{\$OMP SIMD} directive. 
\end{itemize}
Usually, $nblk$ is chosen as a multiple of the vector length. Notice that using a moderate size $nblk$, for the blocks of particles, ensures that the temporary arrays $ww$ and $ll$ fit in cache. 

The SH algorithm  presented on listing \ref{VecSHRoutineCIC} is fully vectorizable and gave very good performances on former Cray machines \cite{sh1987,paruolo1990}. However as we show in the following section, it yields very poor performances on SIMD architectures. 
\begin{lstlisting}[frame=single,caption=Vector version of the charge deposition routine developed by SH for CIC particle shape factors,label=VecSHRoutineCIC, basicstyle=\ttfamily\footnotesize] 
SUBROUTINE depose_rho_vecSH_1_1_1(...)
 ! Declaration and init 
 .....
 ! Loop on particles
  DO ip=1,np,nblk
     !$OMP SIMD
      DO n=ip,MIN(ip+nblk-1,np) !!!! VECTOR
            nn=n-ip+1
             !- Computations relative to particle ip (cell position etc.)
             ...
             ! --- computes weight for each of the 8-vertices of the current cell
             ! --- computes indices of 8-vertices in the array rho
             ind0 = (j+nxguard+1) + (k+nyguard+1)*nnx + (l+nzguard+1)*nnxy
             ww(1,nn) = sx0*sy0*sz0*wq
             ll(1,nn) = ind0+moff(1)
	     ...
	     ...
             ww(8,nn) = sx1*sy1*sz1*wq
             ll(8,nn) = ind0+moff(8)
        END DO
        !$OMP END SIMD
        ! --- add charge density contributions
        DO m= 1,MIN(nblk,np-ip+1)
            !$OMP SIMD
             DO l=1,8  !!!! VECTOR
                 rho(ll(l,m)) = rho(ll(l,m))+ww(l,m)
             END DO
             !$OMP END SIMD
        END DO
  END DO
  ...
END SUBROUTINE depose_rho_vecSH_1_1_1
\end{lstlisting}

\subsubsection{Tests of the  Schwarzmeier and Hewit algorithm on Cori}
\label{TestsCori}
The SH algorithm was tested on one socket of the Cori cluster at NERSC. This socket had one Haswell Xeon processor with the following characteristics: 
\begin{enumerate}[(i)]
\item 16-core CPU at 2.3 GHz, 
\item 256-bit  wide vector unit registers (4 doubles, 8 singles) with AVX2 support, 
\item  256kB L2 cache/core, 40MB shared L3 cache.
\end{enumerate}
The Intel compiler was used to compile the code with option "-O3". The simulation was ran using 1 MPI process and 1 OpenMP thread per MPI process, with the following numerical parameters: 
\begin{enumerate}[(i)]
\item $100\times100\times100$ grid points with $10\times10\times10=1000$ tiles i.e $10$ tiles in each direction, 
\item Two particle species (proton and electron) with 10 particle per cells. The particles are randomly distributed across the simulation domain. The plasma has an initial temperature of $10$~keV. 
\end{enumerate}

The results are displayed on table \ref{testscalarVSvecSH} for order 1 scalar and SH routines, using two different compiler options in each case: 
\begin{enumerate}[(i)]
\item -xCORE-AVX2 to enable vectorization, 
\item -no-vec to disable auto-vectorization of the compiler. In this case, we also manually remove !\$OMP SIMD directives to avoid simd vectorization of loops. 
\end{enumerate}

\begin{table}[h!]
 \centering
 \begin{tabular}{|c|c|c|c|c|}
 \hline
Routine &\multicolumn{2}{|c}{depose\_rho\_scalar\_1\_1\_1} & \multicolumn{2}{|c|}{depose\_rho\_vecSH\_1\_1\_1} \\
\hline
Compiler option &-no-vec & -xCORE-AVX2  & -no-vec  & -xCORE-AVX2 \\
\hline
Time/it/part &$14.6ns$ & $14.6ns$ & $21ns$& $15.9ns$\\
\hline
 \end{tabular}
 \label{testscalarVSvecSH}
 \caption{Performance comparisons of scalar and SH vector routines.   }
 \end{table}

The scalar routine takes the same time for -xCORE-AVX2  and -no-vec options because the routine is not auto-vectorizable by the compiler.

 For the vector routine, we see an improvement of $30\%$ between -xCORE-AVX2  and -no-vec options, showing that vectorization is enabled and working in the -xCORE-AVX2  case. Nevertheless, the overall performance is poor, and the vector routine compiled with -xCORE-AVX2  is even $10\%$ slower than the scalar routine.  

By looking at the code on listing \ref{VecSHRoutineCIC} and using compiler report/ assembly code generated by the Intel compiler, we found two main reasons for this poor performance: 

\begin{enumerate}
\item The first one comes from the strided access of the arrays $ww$ and $ll$ in the loop at line $7$. Assuming cache line sizes of $64$  bytes (8 doubles) and 256-bits wide registers, the four different elements $ww(1,nn_{1})$ to $ww(1,nn_{1}+3)$ are thus on four different cache lines ($ww$ is of size (8,$nblk$)) and this strided access necessitates $4$ stores in memory at different cache lines ("scatter") instead of a single store if the accesses were aligned and contiguous.  A solution would be to switch dimensions of $ww$ but this might not bring any improvement at all because the loop on vertices (line $25$)  would then have strided access for $ww$ ("gather"). Some PIC implementations choose contiguous access for $ww$/$ll$ in the first loop and then use an efficient vector transpose of $ww$/$ll$ before the second loop on vertices. However, this solution requires the use of "shuffle" Intel vector intrinsics to efficiently implement the transpose, which is not portable because this transpose will have to be re-written for a different processor. In addition, this transpose is done $8\times np$ with $np$ the number of particles and might thus add a non-negligible overhead if not done properly.  
\item The second bottleneck comes from the indirect indexing for $rho$ at line $26$. The problem with the current data structure of $rho$ is that the $8$ vertices of one cell are not contiguous in memory, resulting in a rather inefficient gather/scatter instruction.
\end{enumerate}

In the next section, we propose a portable solution for order $1$, $2$ and $3$ charge deposition that solves these two problems and yields a speed-up factor of up to $\times 2.5$ in double precision over the scalar routine. 

\section{New and portable SIMD algorithms}

In this section, we present vector algorithms that perform efficiently on SIMD architectures. 

\subsection{CIC (order $1$) particle shape}

\subsubsection{Algorithm}

The new vector algorithm is detailed on listing \ref{VecHVRoutineCIC}. Similarly to the SH routine, the main particle loop is done by blocks of $nblk$ particles and divided in two consecutive nested loops: $(i)$ a first nested loop that computes particle weights and  $(ii)$ a second one that adds the particle weights to its $8$ nearest vertices.

\subsubsection{Improvements brought by the new algorithm}

 The new algorithm adresses the two main bottlenecks of the SH algorithm with the two following new features: 
\begin{enumerate}
\item a new data structure for $rho$ is introduced, named $rhocells$, which enables memory alignement and unit-stride access when depositing charge on the $8$ vertices. In $rhocells$, the 8-nearest vertices are stored contiguously for each cell.  The array $rhocells$ is thus of size $(8, NCELLS)$ with $NCELLS$ the total number of cells. The element $rhocells(1,icell)$ is therefore 64 bytes-memory aligned for a given cell $icell$ and the elements $rhocells(1:8,icell)$ entirely fit in one cache line allowing for efficient vector load/stores. The array $rhocells$ is reduced to $rho$ once, after the deposition is done for all particles (cf. line $46$). This step is easily vectorizable (see line $48$) but might not lead to optimal performances due to the non-contiguous access in $rho$ that leads to gather-scatter instructions. Notice however that this time, this operation is proportional to the number of cells $NCELLS$ and not to the number of particles $np$ as it was in the case of the SH algorithm. The overhead is thus proportionally lower when there are more particles than cells, which is the case in many PIC simulations of interest, 
\item for each particle, the $8$ different weights $ww$ are now computed using a generic formula (see line $39$) that suppresses  gather instructions formerly needed in the SH algorithm. This also avoids implementing non-portable efficient transpose between the first and second loop, rendering this new algorithm fully portable. 
\end{enumerate}

\begin{lstlisting}[frame=single,caption=New vector version of charge deposition routine  for CIC (order $1$) particle shape factor,label=VecHVRoutineCIC, basicstyle=\ttfamily\footnotesize] 
    SUBROUTINE depose_rho_vecHVv2_1_1_1(...)
        ! Declaration and init
        ...
        nnx = ngridx; nnxy = nnx*ngridy
	moff = (/0,1,nnx,nnx+1,nnxy,nnxy+1,nnxy+nnx,nnxy+nnx+1/)
        mx=(/1_num,0_num,1_num,0_num,1_num,0_num,1_num,0_num/)
        my=(/1_num,1_num,0_num,0_num,1_num,1_num,0_num,0_num/)
        mz=(/1_num,1_num,1_num,1_num,0_num,0_num,0_num,0_num/)
        sgn=(/-1_num,1_num,1_num,-1_num,1_num,-1_num,-1_num,1_num/)

        ! FIRST LOOP: computes cell index of particle and their weight on vertices
        DO ip=1,np,LVEC
            !$OMP SIMD
            DO n=1,MIN(LVEC,np-ip+1)
                nn=ip+n-1
                ! Calculation relative to particle n
                ! --- computes current position in grid units
                x= (xp(nn)-xmin)*dxi
                y = (yp(nn)-ymin)*dyi
                z = (zp(nn)-zmin)*dzi
                ! --- finds cell containing particles for current positions
                j=floor(x)
                k=floor(y)
                l=floor(z)
                ICELL(n)=1+j+nxguard+(k+nyguard+1)*(nx+2*nxguard) &
                +(l+nzguard+1)*(ny+2*nyguard)
                ! --- computes distance between particle and node for current positions
                sx(n) = x-j
                sy(n) = y-k
                sz(n) = z-l
                ! --- computes particles weights
                wq(n)=q*w(nn)*invvol
            END DO
            !$OMP END SIMD
            ! Charge deposition on vertices
            DO n=1,MIN(LVEC,np-ip+1)
                ! --- add charge density contributions to vertices of the current cell
                ic=ICELL(n)
                !$OMP SIMD
                DO nv=1,8 !!! - VECTOR
                    ww=(-mx(nv)+sx(n))*(-my(nv)+sy(n))* &
                        (-mz(nv)+sz(n))*wq(n)*sgn(nv)
                    rhocells(nv,ic)=rhocells(nv,ic)+ww
                END DO
                !$OMP END SIMD
            END DO
        END DO
        ! - reduction of rhocells in rho
	DO iz=1, ncz
            DO iy=1,ncy
                !$OMP SIMD
                DO ix=1,ncx !! VECTOR (take ncx multiple of vector length)
                    ic=ix+(iy-1)*ncx+(iz-1)*ncxy
                    igrid=ic+(iy-1)*ngx+(iz-1)*ngxy
                    rho(orig+igrid+moff(1))=rho(orig+igrid+moff(1))+rhocells(1,ic)
                    rho(orig+igrid+moff(2))=rho(orig+igrid+moff(2))+rhocells(2,ic)
                    rho(orig+igrid+moff(3))=rho(orig+igrid+moff(3))+rhocells(3,ic)
                    rho(orig+igrid+moff(4))=rho(orig+igrid+moff(4))+rhocells(4,ic)
                    rho(orig+igrid+moff(5))=rho(orig+igrid+moff(5))+rhocells(5,ic)
                    rho(orig+igrid+moff(6))=rho(orig+igrid+moff(6))+rhocells(6,ic)
                    rho(orig+igrid+moff(7))=rho(orig+igrid+moff(7))+rhocells(7,ic)
                    rho(orig+igrid+moff(8))=rho(orig+igrid+moff(8))+rhocells(8,ic)
                END DO
                !$OMP END SIMD
            END DO
        END DO

        ...
    END SUBROUTINE depose_rho_vecHVv2_1_1_1
\end{lstlisting}

\subsection{Higher particle shape factors}

Similar algorithms were derived for order $2$ (TSC) and order $3$ particle shape factors, and are detailed in Appendix A. Corresponding current deposition algorithms can be found in  Appendix B for orders $1$, $2$ and $3$ depositions. In these algorithms (see Appenfix B), we use three structures $jxcells$, $jycells$ and $jzcells$ (analogous to $rhocells$ for the deposition of $rho$) for the current components $jx$, $jy$, $jz$ along directions $x$, $y$ and $z$. 

In the following, we detail the data structures used for $rhocells$ for orders $2$ and $3$ particle shapes (cf. Fig. \ref{DataStructVsOrd}): 

\begin{figure}[h!]
\centering
\includegraphics[width=\linewidth]{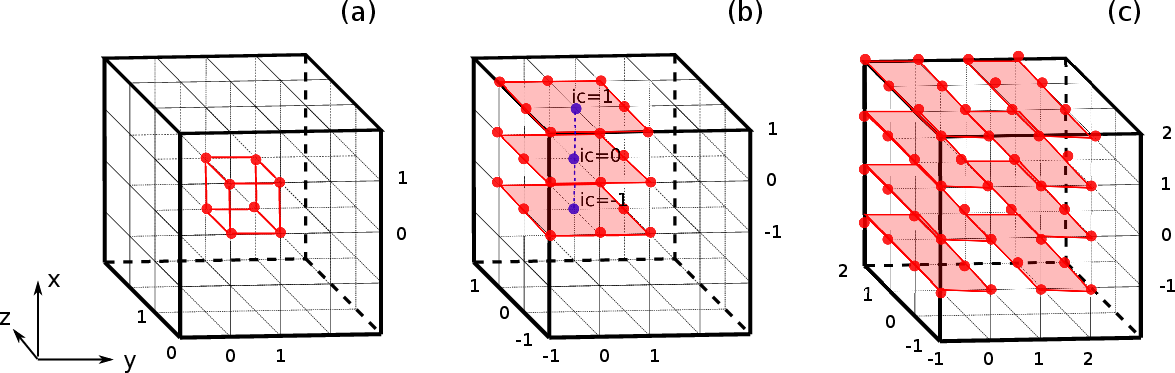}
\caption{\textbf{Data structure used for the array $rhocells$ for different particle shape factors}. In each plot, the particle that deposits charge to its nearest vertices (red/blue points) is located in the cell at position (0,0,0). (a) \textbf{CIC (order $1$) particle shape factor}. The particle deposits its charge to the eight nearest vertices  (red points).   For each cell $icell=(j,k,l)$, $rhocells$ stores the $8$ nearest vertices $(j,k,l)$, $(j+1,k,l)$, $(j,k+1,l)$, $(j+1,k+1,l)$, $(j,k,l+1)$, $(j+1,k,l+1)$, $(j,k+1,l+1)$ and $(j+1,k+1,l+1)$ contiguously. (b) \textbf{TSC (order $2$) particle shape factor}. The particle  deposits its charge to the $27$ neighboring vertices (red and blue points). For a given cell $icell=(j,k,l)$ $rhocells$ stores contiguously the $8$ vertices (red points) $(j,k-1,l-1)$, $(j,k,l-1)$, $(j,k+1,l-1)$, $(j,k-1,l)$, $(j,k+1,l)$, $(j,k-1,l+1)$, $(j,k,l+1)$ and $(j,k+1,l+1)$. The blue points are not stored in rhocells and are treated scalarly in the algorithm. (c) \textbf{QSP (order $3$) particle shape factor}. The particle deposits its charge to the $64$ neighboring vertices (red  points). For a given cell $icell=(j,k,l)$, $rhocells$ stores contiguously the $8$ vertices (delimited by red areas) $(j,k-1,l-1)$, $(j,k,l-1)$, $(j,k+1,l-1)$,  $(j,k+1,l-1)$, $(j,k-1,l)$, $(j,k,l)$, $(j,k+1,l)$,  $(j,k+1,l)$ . }
\label{DataStructVsOrd}
\end{figure}

\begin{enumerate}[(i)]
\item  \textbf{TSC (order $2$) particle shape.} (cf. panel(b) of Fig. \ref{DataStructVsOrd} and listing \ref{VecHVChargeRoutineTSC} in appendix A). In this case, the particles deposit their charge to the $27$ neighbouring vertices. However, storing $27$ contiguous vertices per cell in $rhocells$ would not be efficient as the reduction of $rhocells$ to $rho$ would be much more expensive with potential cache-reuse inefficiency. Instead, while the same size for $rhocells(1:8,1:NCELLS)$ is used, the vertices are now grouped in a different way. The new structure for $rhocells(1:8,1:NCELLS)$ groups $8$ points in a $(y,z)$ plane for each cell $icell$ (see red points in red areas).  For each cell, each particle adds its charge contribution to 24 points in the three planes at $icell-1$, $icell$ and $icell+1$. The three remaining central points (blue points) can be either treated scalarly for $512$-bits wide vector registers or vectorized for $256$-bits by artificially adding a virtual point that does not contribute to any charge. Notice that we did not find a generic formulation for the weights $ww$ and we are therefore still performing a "gather" instruction for $ww$ in the loop on the vertice (line $101$ on listing \ref{VecHVChargeRoutineTSC}). However, this gather is performed in the $y$ and $z$ directions for the first  plane of $8$ points (plane $ic=-1$ on panel (b)) and is subsequently reused on the two other planes $ic=0$ and $ic=1$ (see lines $103$ to $107$ on listing  \ref{VecHVChargeRoutineTSC}). Gather is thus performed only 8 times out of 24 points and thus has a limited impact on performance, as shown below in the reported test results.
 
 \item \textbf{QSP (order $3$) particle shape.} (cf. panel(c) of Fig. \ref{DataStructVsOrd} and listing \ref{VecHVChargeRoutineQSP} in appendix A). In this case, particles deposit their charge to the $64$ neighbouring vertices.  $rhocells(1:8,1:NCELLS)$ also group $8$ points in a ($y$,$z$) plane but differently from the TSC case (see red areas in panel (c)).  For each cell, each particle adds its charge contribution to 64 points in the 8 different $(y,z)$ planes at $icell-ncx-1$, $icell-ncx$, $icell-ncx+1$, $icell-ncx+2$, $icell+ncx-1$,$icell+ncx$,$icell+ncx+1$ and $icell+ncx+2$ where $ncx$ is the number of cells in the $x$ direction (see lines $63$ to $77$ on listing \ref{VecHVChargeRoutineQSP}). This might reduce the flop/byte ratio of the second loop  when $nnx$ is large enough so that elements $rhocells(1:8,icell)$ and $rhocells(1:8,icell+nnx-1)$ are not in $L1$ cache. The vertices could have been grouped in $(y,z)$ planes of $16$ points instead of $8$ points but this would imply a bigger reduction loop of $rhocells$ in $rho$ and worst performances for a low number of particles. Notice that here again, we did not find an efficient generic formulation for the weights $ww$ and we are therefore still performing a "gather" instruction  (see lines $116$ and $126$ on listing \ref{VecHVChargeRoutineQSP}). However, this gather is performed in the $y$ and $z$ directions and is subsequently for computing the weights at different positions in $x$  (see lines $118$ to $124$ and $128$ to $134$ on listing  \ref{VecHVChargeRoutineQSP}). Gather is thus performed only 16 times out of 64 points and thus has a limited impact on performance, as shown below in the reported test results.

\end{enumerate}
\section{Benchmarks of the new algorithms}

\begin{figure}[h!]
\centering
\includegraphics[width=\linewidth]{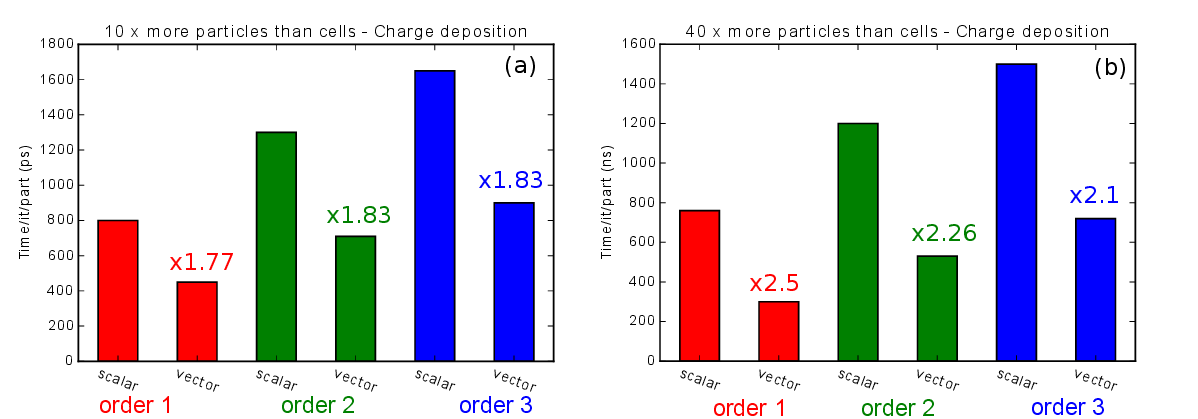}
\label{BenchmarksChargeCori}
\caption{\textbf{Benchmarks of the new $3D$ charge deposition algorithms on Cori.} Each bar plot shows the time/it/part in $ps$ for different particle shape orders $1$ to $3$.  (a) Benchmarks with $10$ times more particles than cells. (b) Benchmarks with $40$ times more particles than cells.  }
\end{figure}

\begin{figure}[h!]
\centering
\includegraphics[width=\linewidth]{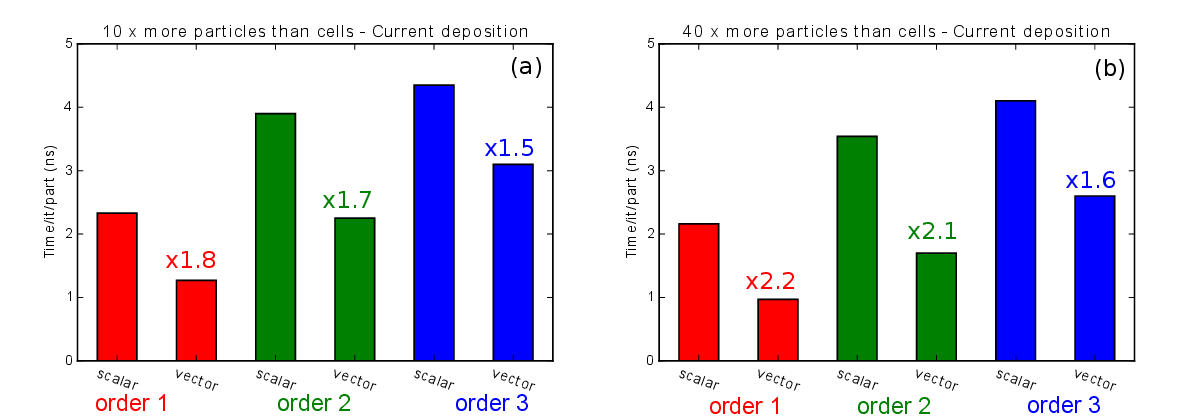}
\label{BenchmarksCurrentCori}
\caption{\textbf{Benchmarks of the new $3D$ current deposition algorithms on Cori.} Each bar plot shows the time/it/part in $ns$ for different particle shape orders $1$ to $3$.  (a) Benchmarks with $10$ times more particles than cells. (b) Benchmarks with $40$ times more particles than cells.  }
\end{figure}

The new vector algorithms were benchmarked on one node (two sockets) of the Cori machine in the same numerical conditions than the ones used in section \ref{TestsCori} but with 2 MPI processes (one per socket) and 16 OpenMP threads per MPI process. For charge deposition, we use 10x10x10 tiles in each direction. For current deposition, we use a larger number of tiles (12x12x12 tiles in each direction) so that the three structures $jxcells$, $jycells$ and $jzcells$ (equivalent of $rhocells$ for current deposition) fit in cache.  Results are shown on Fig.  \ref{BenchmarksChargeCori} for charge deposition and  on Fig.  \ref{BenchmarksCurrentCori} for current deposition. Panels (a) show the time/iteration/particle (in $ps$ for Fig.  \ref{BenchmarksChargeCori} and ns for Fig.  \ref{BenchmarksCurrentCori} ) taken by the  deposition routines for different particle shape factors and when there are $10$ times more particles than cells. Panels (b) show the same quantities but for $40$ times more particles than cells. 

Notice that as we vectorize on vertices, there is no performance bottleneck related to a possibly inhomogeneous distribution of particles on the simulation domain. Even for a low number of particles per cell (e.g panel (a) of Fig.  \ref{BenchmarksChargeCori}), the algorithm performs well, with speed-ups of up to $\times 1.8$. When the number of particles increases (Fig.  \ref{BenchmarksChargeCori} of panel (b)) performances are even better because the reduction operation of $rhocells$ in $rho$ becomes more and more negligible relatively to particle loops. For $40$ times more particles than cells, performances now reach $\times 2.5$ for order $1$ particle shape factor. Order $3$ deposition performs less efficiently than orders $1$ and $2$, because as we described in the previous section, the structure we chose for $rhocells$ decreases the flop/byte ratio of the loop on vertices compared to orders $1$ and $2$. In the case of simulations using a lot of particles, for which the reduction of $rhocells$ in $rho$ is negligible, one might consider grouping vertices in $rhocells$ by groups of $16$ instead of $8$ for order $3$ deposition in order to increase the flop/byte ratio in loop on vertices.

\section{Conclusion and prospects}

A new method is presented that allows for efficient vectorization of the standard charge/current deposition routines on current SIMD architectures, leading to efficient deposition algorithms for shape factors of order 1, 2 and 3. The algorithms can be used on current multi-core architectures (with up to AVX2 support) as well as on future many-core Intel $KNL$ processors that will support $AVX-512$. Further tests on KNL will be performed as the processor becomes available. 

This work provides  deposition routines that are fully portable and only use the \textit{\$OMP SIMD} directives that are provided by OpenMP $4.0$. Efficient vectorization of the charge conserving current deposition from Esirkepov is being investigated, and will be detailed in future work. 

 \section*{Acknowledgement}

We thank Karthik Raman from Intel inc for useful discussions on the optimization of the vector routines. This work was supported by the European Commission through the Marie Sk\l owdoska-Curie actions (Marie Curie IOF fellowship PICSSAR grant number 624543) as well as by the Director, Office of Science, Office of High Energy Physics, U.S. Dept. of Energy under Contract No. DE-AC02-05CH11231, the US-DOE SciDAC program ComPASS, and the US-DOE program CAMPA. This research used resources of the National Energy Research Scientific Computing Center, a DOE Office of Science User Facility supported by the Office of Science of the U.S. Department of Energy under Contract No. DE-AC02-05CH11231.

This document was prepared as an account of work sponsored in part
by the United States Government. While this document is believed to
contain correct information, neither the United States Government
nor any agency thereof, nor The Regents of the University of California,
nor any of their employees, nor the authors makes any warranty, express
or implied, or assumes any legal responsibility for the accuracy,
completeness, or usefulness of any information, apparatus, product,
or process disclosed, or represents that its use would not infringe
privately owned rights. Reference herein to any specific commercial
product, process, or service by its trade name, trademark, manufacturer,
or otherwise, does not necessarily constitute or imply its endorsement,
recommendation, or favoring by the United States Government or any
agency thereof, or The Regents of the University of California. The
views and opinions of authors expressed herein do not necessarily
state or reflect those of the United States Government or any agency
thereof or The Regents of the University of California.

\newpage

\bibliographystyle{ieeetr}

\newpage
\appendix 

\section{Full vector algorithms in Fortran 90 for order $1$, $2$ and $3$ charge deposition routines }
In the following we use the notations below for input/output parameters of charge deposition subroutines: 

\begin{itemize}
\item $rho$ is the charge density (grid array),  
\item $np$ is the number of particles (scalar),  
\item $xp,yp,zp$ are particle positions (particle arrays)
\item  $w$ is the particle weights (particle array) and $q$ the particle species charge (scalar)
\item $xmin,ymin,zmin$ are the absolute coordinates (scalars) of the origin of the current spatial partition (tile or MPI subdomain depending on implementation) containing particle arrays (tile or subdomain), 
\item $dx,dy,dz$ (scalars) are the spatial mesh size in each direction, 
\item $nx,ny,nz$ (scalars) are the number of cells in each direction (without guard cells) of the current spatial partition, 
\item $nxguard,nyguard,nzguard$ (scalars) are the number of guard cells in each direction of the current spatial partition.  
\end{itemize}

\subsection{Order $1$ charge deposition routine}
\begin{lstlisting}[frame=single,caption=New vector version of charge deposition routine  for CIC particle shape factors,label=VecHVChargeRoutineCIC, basicstyle=\ttfamily\footnotesize] 
    SUBROUTINE depose_rho_vecHVv2_1_1_1(rho,np,xp,yp,zp,w,q,xmin,ymin,zmin, &
    dx,dy,dz,nx,ny,nz,nxguard,nyguard,nzguard)
        USE constants
        IMPLICIT NONE
        INTEGER, INTENT (IN) :: np,nx,ny,nz,nxguard,nyguard,nzguard
        REAL(num),INTENT(IN OUT) :: rho(1:(1+nx+2*nxguard)* &
        (1+ny+2*nyguard)*(1+nz+2*nzguard))
        REAL(num), DIMENSION(:,:), ALLOCATABLE:: rhocells
        INTEGER, PARAMETER :: LVEC=64
        INTEGER, DIMENSION(LVEC) :: ICELL
        REAL(num) :: ww
        INTEGER :: NCELLS
        REAL(num) :: xp(np), yp(np), zp(np), w(np)
        REAL(num) :: q,dt,dx,dy,dz,xmin,ymin,zmin
        REAL(num) :: dxi,dyi,dzi
        REAL(num) :: xint,yint,zint
        REAL(num) :: x,y,z,invvol
        REAL(num) :: sx(LVEC), sy(LVEC), sz(LVEC), wq(LVEC)
        REAL(num), PARAMETER :: onesixth=1.0_num/6.0_num,twothird=2.0_num/3.0_num
        INTEGER :: ic,igrid,j,k,l,vv,n,ip,jj,kk,ll,nv,nn
        INTEGER :: nnx, nnxy
        INTEGER :: moff(1:8) 
        REAL(num):: mx(1:8),my(1:8),mz(1:8), sgn(1:8)
        INTEGER :: orig, jorig, korig, lorig
        INTEGER :: ncx, ncy, ncxy, ncz,ix,iy,iz, ngridx, ngridy, ngx, ngxy

        ! Init parameters
        dxi = 1.0_num/dx
        dyi = 1.0_num/dy
        dzi = 1.0_num/dz
        invvol = dxi*dyi*dzi
        ngridx=nx+1+2*nxguard;ngridy=ny+1+2*nyguard;
        ncx=nx+2;ncy=ny+2;ncz=nz+2
        NCELLS=ncx*ncy*ncz
        ALLOCATE(rhocells(8,NCELLS))
        rhocells=0.0_num
        nnx = ngridx
        nnxy = nnx*ngridy
        moff = (/0,1,nnx,nnx+1,nnxy,nnxy+1,nnxy+nnx,nnxy+nnx+1/)
        mx=(/1_num,0_num,1_num,0_num,1_num,0_num,1_num,0_num/)
        my=(/1_num,1_num,0_num,0_num,1_num,1_num,0_num,0_num/)
        mz=(/1_num,1_num,1_num,1_num,0_num,0_num,0_num,0_num/)
        sgn=(/-1_num,1_num,1_num,-1_num,1_num,-1_num,-1_num,1_num/)
        jorig=-1; korig=-1;lorig=-1
        orig=jorig+nxguard+nnx*(korig+nyguard)+(lorig+nzguard)*nnxy
        ngx=(ngridx-ncx)
        ngxy=(ngridx*ngridy-ncx*ncy)
        ncxy=ncx*ncy
        ! FIRST LOOP: computes cell index of particle and their weight on vertices
        DO ip=1,np,LVEC
            !$OMP SIMD
            DO n=1,MIN(LVEC,np-ip+1)
                nn=ip+n-1
                ! Calculation relative to particle n
                ! --- computes current position in grid units
                x= (xp(nn)-xmin)*dxi
                y = (yp(nn)-ymin)*dyi
                z = (zp(nn)-zmin)*dzi
                ! --- finds cell containing particles for current positions
                j=floor(x)
                k=floor(y)
                l=floor(z)
                ICELL(n)=1+(j-jorig)+(k-korig)*(ncx)+(l-lorig)*ncxy
                ! --- computes distance between particle and node for current positions
                sx(n) = x-j
                sy(n) = y-k
                sz(n) = z-l
                ! --- computes particles weights
                wq(n)=q*w(nn)*invvol
            END DO
            !$OMP END SIMD
            ! Current deposition on vertices
            DO n=1,MIN(LVEC,np-ip+1)
                ! --- add charge density contributions to vertices of the current cell
                ic=ICELL(n)
                !$OMP SIMD
                DO nv=1,8 !!! - VECTOR
                    ww=(-mx(nv)+sx(n))*(-my(nv)+sy(n))* &
                        (-mz(nv)+sz(n))*wq(n)*sgn(nv)
                    rhocells(nv,ic)=rhocells(nv,ic)+ww
                END DO
                !$OMP END SIMD
            END DO
        END DO
        ! - reduction of rhocells in rho
        DO iz=1, ncz
            DO iy=1,ncy
                !$OMP SIMD
                DO ix=1,ncx !! VECTOR (take ncx multiple of vector length)
                    ic=ix+(iy-1)*ncx+(iz-1)*ncxy
                    igrid=ic+(iy-1)*ngx+(iz-1)*ngxy
                    rho(orig+igrid+moff(1))=rho(orig+igrid+moff(1))+rhocells(1,ic)
                    rho(orig+igrid+moff(2))=rho(orig+igrid+moff(2))+rhocells(2,ic)
                    rho(orig+igrid+moff(3))=rho(orig+igrid+moff(3))+rhocells(3,ic)
                    rho(orig+igrid+moff(4))=rho(orig+igrid+moff(4))+rhocells(4,ic)
                    rho(orig+igrid+moff(5))=rho(orig+igrid+moff(5))+rhocells(5,ic)
                    rho(orig+igrid+moff(6))=rho(orig+igrid+moff(6))+rhocells(6,ic)
                    rho(orig+igrid+moff(7))=rho(orig+igrid+moff(7))+rhocells(7,ic)
                    rho(orig+igrid+moff(8))=rho(orig+igrid+moff(8))+rhocells(8,ic)
                END DO
                !$OMP END SIMD
            END DO
        END DO
        DEALLOCATE(rhocells)
        RETURN
    END SUBROUTINE depose_rho_vecHVv2_1_1_1
\end{lstlisting}
\subsection{Order $2$ charge deposition routine}

\begin{lstlisting}[frame=single,caption=New vector version of charge deposition routine  for TSC particle shape factors,label=VecHVChargeRoutineTSC, basicstyle=\ttfamily\footnotesize] 
SUBROUTINE depose_rho_vecHVv2_2_2_2(rho,np,xp,yp,zp,w,q,xmin,ymin,zmin, &
	dx,dy,dz,nx,ny,nz,nxguard,nyguard,nzguard)
        USE constants
        IMPLICIT NONE
        INTEGER :: np,nx,ny,nz,nxguard,nyguard,nzguard
        REAL(num),INTENT(IN OUT) :: rho(1:(1+nx+2*nxguard)* &
        (1+ny+2*nyguard)*(1+nz+2*nzguard))
        REAL(num), DIMENSION(:,:), ALLOCATABLE:: rhocells
        INTEGER, PARAMETER :: LVEC=64
        INTEGER, DIMENSION(LVEC) :: ICELL, IG
        REAL(num) :: ww, wwx,wwy,wwz
        INTEGER :: NCELLS
        REAL(num) :: xp(np), yp(np), zp(np), w(np)
        REAL(num) :: q,dt,dx,dy,dz,xmin,ymin,zmin
        REAL(num) :: dxi,dyi,dzi
        REAL(num) :: xint,yint,zint,xintsq,yintsq,zintsq
        REAL(num) :: x,y,z,invvol, wq0, wq, szy, syy0,syy1,syy2,szz0,szz1,szz2
        REAL(num) :: sx0(LVEC), sx1(LVEC), sx2(LVEC)
        REAL(num), PARAMETER :: onesixth=1.0_num/6.0_num,twothird=2.0_num/3.0_num
        INTEGER :: ic,igrid,j,k,l,vv,n,ip,jj,kk,ll,nv,nn
        INTEGER :: nnx, nnxy, off0, ind0
        INTEGER :: moff(1:8)
        REAL(num):: ww0(1:LVEC,1:8),www(1:LVEC,1:8)
        INTEGER :: orig, jorig, korig, lorig
        INTEGER :: ncx, ncy, ncxy, ncz,ix,iy,iz, ngridx, ngridy, ngx, ngxy

        ! Init parameters
        dxi = 1.0_num/dx
        dyi = 1.0_num/dy
        dzi = 1.0_num/dz
        invvol = dxi*dyi*dzi
        wq0=q*invvol
        ngridx=nx+1+2*nxguard;ngridy=ny+1+2*nyguard
        ncx=nx+3;ncy=ny+3;ncz=nz+3
        NCELLS=ncx*ncy*ncz
        ALLOCATE(rhocells(8,NCELLS))
        rhocells=0.0_num
        nnx = nx + 1 + 2*nxguard
        nnxy = nnx*(ny+1+2*nyguard)
        moff = (/-nnx-nnxy,-nnxy,nnx-nnxy,-nnx,nnx,-nnx+nnxy,nnxy,nnx+nnxy/)
	    ww0=0.0_num
        jorig=-1; korig=-1;lorig=-1
        orig=jorig+nxguard+nnx*(korig+nyguard)+(lorig+nzguard)*nnxy
        ngx=(ngridx-ncx)
        ngxy=(ngridx*ngridy-ncx*ncy)
        ncxy=ncx*ncy
        ! FIRST LOOP: computes cell index of particle and their weight on vertices
        DO ip=1,np,LVEC
            !$OMP SIMD
            DO n=1,MIN(LVEC,np-ip+1)
                nn=ip+n-1
                ! Calculation relative to particle n
                ! --- computes current position in grid units
                x= (xp(nn)-xmin)*dxi
                y = (yp(nn)-ymin)*dyi
                z = (zp(nn)-zmin)*dzi
                ! --- finds cell containing particles for current positions
                j=nint(x)
                k=nint(y)
                l=nint(z)
                ICELL(n)=1+(j-jorig)+(k-korig)*(ncx)+(l-lorig)*ncxy
                IG(n)=ICELL(n)+(k-korig)*ngx+(l-lorig)*ngxy
                ! --- computes distance between particle and node for current positions
                xint = x-j
                yint = y-k
                zint = z-l
                xintsq=xint**2
                yintsq=yint**2
                zintsq=zint**2
                ! --- computes particles weights
                wq=w(nn)*wq0
                sx0(n)=0.5_num*(0.5_num-xint)**2
                sx1(n)=(0.75_num-xintsq)
                sx2(n)=0.5_num*(0.5_num+xint)**2
                syy0=0.5_num*(0.5_num-yint)**2
                syy1=(0.75_num-yintsq)
                syy2=0.5_num*(0.5_num+yint)**2
                szz0=0.5_num*(0.5_num-zint)**2*wq
                szz1=(0.75_num-zintsq)*wq
                szz2=0.5_num*(0.5_num+zint)**2*wq
                www(n,1) = syy0*szz0
                www(n,2) = syy1*szz0
                www(n,3) = syy2*szz0
                www(n,4) = syy0*szz1
                www(n,5) = syy2*szz1
                www(n,6) = syy0*szz2
                www(n,7) = syy1*szz2
                www(n,8) = syy2*szz2
                szy=syy1*szz1 ! central point
                ww0(n,1)=szy*sx0(n)
                ww0(n,2)=szy*sx1(n)
                ww0(n,3)=szy*sx2(n)
            END DO
            !$OMP END SIMD
            ! Current deposition on vertices
            DO n=1,MIN(LVEC,np-ip+1)
                ! --- add charge density contributions to vertices of the current cell
                !DIR$ ASSUME_ALIGNED rhocells:64
                !$OMP SIMD
                DO nv=1,8 !!! - VECTOR
                    ww=www(n,nv)
                    ! Loop on (i=-1,j,k)
                    rhocells(nv,ICELL(n)-1)=rhocells(nv,ICELL(n)-1)+ww*sx0(n)
                    ! Loop on (i=0,j,k)
                    rhocells(nv,ICELL(n))=rhocells(nv,ICELL(n))+ww*sx1(n)
                    !Loop on (i=1,j,k)
                    rhocells(nv,ICELL(n)+1)=rhocells(nv,ICELL(n)+1)+ww*sx2(n)
                END DO
                !$OMP END SIMD
                !$OMP SIMD
                DO nv=1,4
                    rho(orig+IG(n)+nv-2)=rho(orig+IG(n)+nv-2)+ww0(n,nv)
                END DO
                !$OMP END SIMD
            END DO
        END DO
        ! - reduction of rhocells in rho
        DO iz=1, ncz
            DO iy=1,ncy
                !$OMP SIMD
                DO ix=1,ncx !! VECTOR (take ncx multiple of vector length)
                    ic=ix+(iy-1)*ncx+(iz-1)*ncxy
                    igrid=ic+(iy-1)*ngx+(iz-1)*ngxy
                    rho(orig+igrid+moff(1))=rho(orig+igrid+moff(1))+rhocells(1,ic)
                    rho(orig+igrid+moff(2))=rho(orig+igrid+moff(2))+rhocells(2,ic)
                    rho(orig+igrid+moff(3))=rho(orig+igrid+moff(3))+rhocells(3,ic)
                    rho(orig+igrid+moff(4))=rho(orig+igrid+moff(4))+rhocells(4,ic)
                    rho(orig+igrid+moff(5))=rho(orig+igrid+moff(5))+rhocells(5,ic)
                    rho(orig+igrid+moff(6))=rho(orig+igrid+moff(6))+rhocells(6,ic)
                    rho(orig+igrid+moff(7))=rho(orig+igrid+moff(7))+rhocells(7,ic)
                    rho(orig+igrid+moff(8))=rho(orig+igrid+moff(8))+rhocells(8,ic)
                END DO
                !$OMP END SIMD
            END DO
        END DO
        DEALLOCATE(rhocells)
        RETURN
    END SUBROUTINE depose_rho_vecHVv2_2_2_2

\end{lstlisting}

\subsection{Order $3$ charge deposition routine}

\begin{lstlisting}[frame=single,caption=Vector version of charge deposition routine developed by SH for QSP particle shape factors,label=VecHVChargeRoutineQSP, basicstyle=\ttfamily\footnotesize] 
SUBROUTINE depose_rho_vecHVv2_3_3_3(rho,np,xp,yp,zp,w,q,xmin,ymin,zmin, &
	dx,dy,dz,nx,ny,nz,nxguard,nyguard,nzguard)
        USE constants
        IMPLICIT NONE
        INTEGER :: np,nx,ny,nz,nxguard,nyguard,nzguard
        REAL(num),INTENT(IN OUT) :: rho(1:(1+nx+2*nxguard)* &
        (1+ny+2*nyguard)*(1+nz+2*nzguard))
        REAL(num), DIMENSION(:,:), ALLOCATABLE:: rhocells
        INTEGER, PARAMETER :: LVEC=16
        INTEGER, DIMENSION(LVEC) :: ICELL
        REAL(num) :: ww, wwx,wwy,wwz
        INTEGER :: NCELLS
        REAL(num) :: xp(np), yp(np), zp(np), w(np)
        REAL(num) :: q,dt,dx,dy,dz,xmin,ymin,zmin
        REAL(num) :: dxi,dyi,dzi,xint,yint,zint(1:LVEC), &
                   oxint,oyint,ozint,xintsq,yintsq,zintsq,oxintsq,oyintsq,ozintsq
        REAL(num) :: x,y,z,invvol, wq0, wq
        REAL(num) :: sx1(LVEC), sx2(LVEC), sx3(LVEC),sx4(LVEC), sy1, sy2, sy3,sy4, &
                     sz1, sz2, sz3,sz4, w1,w2
        REAL(num), PARAMETER :: onesixth=1.0_num/6.0_num,twothird=2.0_num/3.0_num
        INTEGER :: ic, igrid, ic0,j,k,l,vv,n,ip,jj,kk,ll,nv,nn
        INTEGER :: nnx, nnxy, off0, ind0
        INTEGER :: moff(1:8)
        REAL(num):: www1(LVEC,8),www2(LVEC,8), zdec(1:8), &
        h1(1:8), h11(1:8), h12(1:8), sgn(1:8), szz(1:8)
        INTEGER :: orig, jorig, korig, lorig
        INTEGER :: ncx, ncy, ncxy, ncz,ix,iy,iz, ngridx, ngridy, ngx, ngxy

        ! Init parameters
        dxi = 1.0_num/dx
        dyi = 1.0_num/dy
        dzi = 1.0_num/dz
        invvol = dxi*dyi*dzi
        wq0=q*invvol
        ngridx=nx+1+2*nxguard;ngridy=ny+1+2*nyguard
        ncx=nx+5; ncy=ny+4; ncz=nz+2
        NCELLS=ncx*ncy*ncz
        ALLOCATE(rhocells(8,NCELLS))
        rhocells=0_num
        nnx = ngridx
        nnxy = ngridx*ngridy
        moff = (/-nnxy,0,nnxy,2*nnxy,nnx-nnxy,nnx,nnx+nnxy,nnx+2*nnxy/)
        jorig=-2; korig=-2;lorig=-1
        orig=jorig+nxguard+nnx*(korig+nyguard)+(lorig+nzguard)*nnxy
        ngx=(ngridx-ncx)
        ngxy=(ngridx*ngridy-ncx*ncy)
        ncxy=ncx*ncy

        ! FIRST LOOP: computes cell index of particle and their weight on vertices
        DO ip=1,np,LVEC
            !$OMP SIMD
            DO n=1,MIN(LVEC,np-ip+1)
                nn=ip+n-1
                ! Calculation relative to particle n
                ! --- computes current position in grid units
                x= (xp(nn)-xmin)*dxi
                y = (yp(nn)-ymin)*dyi
                z = (zp(nn)-zmin)*dzi
                ! --- finds cell containing particles for current positions
                j=floor(x)
                k=floor(y)
                l=floor(z)
                ICELL(n)=1+(j-jorig)+(k-korig)*(ncx)+(l-lorig)*ncxy
                wq=w(nn)*wq0
                ! --- computes distance between particle and node for current positions
                xint = x-j
                yint= y-k
                zint(n) = z-l
                ! --- computes coefficients for node centered quantities
                oxint = 1.0_num-xint
                xintsq = xint*xint
                oxintsq = oxint*oxint
                sx1(n) = onesixth*oxintsq*oxint
                sx2(n) = twothird-xintsq*(1.0_num-xint*0.5_num)
                sx3(n) = twothird-oxintsq*(1.0_num-oxint*0.5_num)
                sx4(n) = onesixth*xintsq*xint
                oyint = 1.0_num-yint
                yintsq = yint*yint
                oyintsq = oyint*oyint
                sy1 = onesixth*oyintsq*oyint
                sy2 = (twothird-yintsq*(1.0_num-yint*0.5_num))
                sy3 = (twothird-oyintsq*(1.0_num-oyint*0.5_num))
                sy4 = onesixth*yintsq*yint
                ozint = 1.0_num-zint(n)
                zintsq = zint(n)*zint(n)
                ozintsq = ozint*ozint
                sz1 = onesixth*ozintsq*ozint*wq
                sz2 = (twothird-zintsq*(1.0_num-zint(n)*0.5_num))*wq
                sz3 = (twothird-ozintsq*(1.0_num-ozint*0.5_num))*wq
                sz4 = onesixth*zintsq*zint(n)*wq
                www1(n,1)=sz1*sy1
                www1(n,2)=sz2*sy1
                www1(n,3)=sz3*sy1
                www1(n,4)=sz4*sy1
                www1(n,5)=sz1*sy2
                www1(n,6)=sz2*sy2
                www1(n,7)=sz3*sy2
                www1(n,8)=sz4*sy2
                www2(n,1)=sz1*sy3
                www2(n,2)=sz2*sy3
                www2(n,3)=sz3*sy3
                www2(n,4)=sz4*sy3
                www2(n,5)=sz1*sy4
                www2(n,6)=sz2*sy4
                www2(n,7)=sz3*sy4
                www2(n,8)=sz4*sy4
            END DO
            !$OMP END SIMD
            ! Current deposition on vertices
            DO n=1,MIN(LVEC,np-ip+1)
                ! --- add charge density contributions to vertices of the current cell
                ic=ICELL(n)
                !DIR$ ASSUME_ALIGNED rhocells:64, www1:64, www2:64
                !$OMP SIMD
                DO nv=1,8 !!! - VECTOR
                    w1=www1(n,nv)
                    ! Loop on (i=-1,j,k)
                    rhocells(nv,ic-ncx-1) = rhocells(nv,ic-ncx-1) + w1*sx1(n)
                    ! Loop on (i=0,j,k)
                    rhocells(nv,ic-ncx)   = rhocells(nv,ic-ncx)   + w1*sx2(n)
                    !Loop on (i=1,j,k)
                    rhocells(nv,ic-ncx+1) = rhocells(nv,ic-ncx+1) + w1*sx3(n)
                    !Loop on (i=1,j,k)
                    rhocells(nv,ic-ncx+2) = rhocells(nv,ic-ncx+2) + w1*sx4(n)

                    w2=www2(n,nv)
                    ! Loop on (i=-1,j,k)
                    rhocells(nv,ic+ncx-1) = rhocells(nv,ic+ncx-1) + w2*sx1(n)
                    ! Loop on (i=0,j,k)
                    rhocells(nv,ic+ncx)   = rhocells(nv,ic+ncx)   + w2*sx2(n)
                    !Loop on (i=1,j,k)
                    rhocells(nv,ic+ncx+1) = rhocells(nv,ic+ncx+1) + w2*sx3(n)
                    !Loop on (i=1,j,k)
                    rhocells(nv,ic+ncx+2) = rhocells(nv,ic+ncx+2) + w2*sx4(n)
                END DO
                !$OMP END SIMD
            END DO
        END DO
        ! - reduction of rhocells in rho
        DO iz=1, ncz
            DO iy=1,ncy
                !$OMP SIMD
                DO ix=1,ncx !! VECTOR (take ncx multiple of vector length)
                    ic=ix+(iy-1)*ncx+(iz-1)*ncxy
                    igrid=ic+(iy-1)*ngx+(iz-1)*ngxy
                    rho(orig+igrid+moff(1))=rho(orig+igrid+moff(1))+rhocells(1,ic)
                    rho(orig+igrid+moff(2))=rho(orig+igrid+moff(2))+rhocells(2,ic)
                    rho(orig+igrid+moff(3))=rho(orig+igrid+moff(3))+rhocells(3,ic)
                    rho(orig+igrid+moff(4))=rho(orig+igrid+moff(4))+rhocells(4,ic)
                    rho(orig+igrid+moff(5))=rho(orig+igrid+moff(5))+rhocells(5,ic)
                    rho(orig+igrid+moff(6))=rho(orig+igrid+moff(6))+rhocells(6,ic)
                    rho(orig+igrid+moff(7))=rho(orig+igrid+moff(7))+rhocells(7,ic)
                    rho(orig+igrid+moff(8))=rho(orig+igrid+moff(8))+rhocells(8,ic)
                END DO
                !$OMP END SIMD
            END DO
        END DO
        DEALLOCATE(rhocells)
        RETURN
    END SUBROUTINE depose_rho_vecHVv2_3_3_3

    \end{lstlisting}

\section{Full vector algorithms in Fortran 90 for order $1$, $2$ and $3$ current deposition routines }    
 
 In the following we use the notations below for input/output parameters of charge deposition subroutines: 

\begin{itemize}
\item $jx$, $jy$, $jz$ are the currents in $x$, $y$, $z$ (grid array),  
\item $np$ is the number of particles (scalar),  
\item $xp,yp,zp$ are particle positions (particle arrays)
\item  $w$ is the particle weights (particle array) and $q$ the particle species charge (scalar)
\item $xmin,ymin,zmin$ are the absolute coordinates (scalars) of the origin of the current spatial partition (tile or MPI subdomain depending on implementation) containing particle arrays (tile or subdomain), 
\item $dx,dy,dz$ (scalars) are the spatial mesh size in each direction, 
\item $nx,ny,nz$ (scalars) are the number of cells in each direction (without guard cells) of the current spatial partition, 
\item $nxguard,nyguard,nzguard$ (scalars) are the number of guard cells in each direction of the current spatial partition.  
\end{itemize}

\subsection{Order $1$ current deposition routine}
\begin{lstlisting}[frame=single,caption=New vector version of current deposition routine  for CIC particle shape factors,label=VecHVCurrentRoutineCIC, basicstyle=\ttfamily\footnotesize] 
SUBROUTINE depose_jxjyjz_vecHVv2_1_1_1(jx,jy,jz,np,xp,yp,zp,uxp,uyp,uzp,w,q, &
    xmin,ymin,zmin,dt,dx,dy,dz,nx,ny,nz,nxguard,nyguard,nzguard)
    USE constants
    IMPLICIT NONE
    INTEGER :: np,nx,ny,nz,nxguard,nyguard,nzguard
    REAL(num),INTENT(IN OUT) :: jx(1:(1+nx+2*nxguard)*(1+ny+2*nyguard)*(1+nz+2*nzguard))
    REAL(num),INTENT(IN OUT) :: jy(1:(1+nx+2*nxguard)*(1+ny+2*nyguard)*(1+nz+2*nzguard))
    REAL(num),INTENT(IN OUT) :: jz(1:(1+nx+2*nxguard)*(1+ny+2*nyguard)*(1+nz+2*nzguard))
    REAL(num), DIMENSION(:,:), ALLOCATABLE:: jxcells,jycells,jzcells
    REAL(num), DIMENSION(np) :: xp,yp,zp,uxp,uyp,uzp, w
    REAL(num) :: q,dt,dx,dy,dz,xmin,ymin,zmin
    REAL(num) :: dxi,dyi,dzi,xint,yint,zint, &
                   oxint,oyint,ozint,xintsq,yintsq,zintsq,oxintsq,oyintsq,ozintsq
    REAL(num) :: x,y,z,xmid,ymid,zmid,invvol, dts2dx, dts2dy, dts2dz
    REAL(num) ::  gaminv, usq, clightsq
    REAL(num), PARAMETER :: onesixth=1.0_num/6.0_num,twothird=2.0_num/3.0_num
    INTEGER :: j,k,l,j0,k0,l0,ip, NCELLS, ic
    INTEGER :: nnx, nnxy, n,nn,nv
    INTEGER :: moff(1:8) 
    REAL(num):: mx(1:8),my(1:8),mz(1:8), sgn(1:8)
    INTEGER, PARAMETER :: LVEC=8
    INTEGER, DIMENSION(LVEC,3) :: ICELL
    REAL(num), DIMENSION(LVEC) :: sx, sy, sz, sx0, sy0, sz0,wqx,wqy,wqz
    REAL(num) :: wwx,wwy,wwz, wq,vx,vy,vz, wx,wx0, wy,wy0, wz,wz0
    INTEGER :: orig, jorig, korig, lorig, igrid
    INTEGER :: ncx, ncy, ncxy, ncz,ix,iy,iz, ngridx, ngridy, ngx, ngxy

    dxi = 1.0_num/dx
    dyi = 1.0_num/dy
    dzi = 1.0_num/dz
    invvol = dxi*dyi*dzi
    dts2dx = 0.5_num*dt*dxi
    dts2dy = 0.5_num*dt*dyi
    dts2dz = 0.5_num*dt*dzi
    clightsq = 1.0_num/clight**2
    sx=0.0_num;sy=0.0_num;sz=0.0_num
    sx0=0.0_num;sy0=0.0_num;sz0=0.0_num
    ngridx=nx+1+2*nxguard;ngridy=ny+1+2*nyguard;
    ncx=nx+3;ncy=ny+3;ncz=nz+3
    NCELLS=ncx*ncy*ncz
    ALLOCATE(jxcells(8,NCELLS),jycells(8,NCELLS),jzcells(8,NCELLS))
    jxcells=0.0_num; jycells=0.0_num; jzcells=0.0_num;
    nnx = ngridx
    nnxy = nnx*ngridy
    moff = (/0,1,nnx,nnx+1,nnxy,nnxy+1,nnxy+nnx,nnxy+nnx+1/)
    mx=(/1_num,0_num,1_num,0_num,1_num,0_num,1_num,0_num/)
    my=(/1_num,1_num,0_num,0_num,1_num,1_num,0_num,0_num/)
    mz=(/1_num,1_num,1_num,1_num,0_num,0_num,0_num,0_num/)
    sgn=(/-1_num,1_num,1_num,-1_num,1_num,-1_num,-1_num,1_num/)
    jorig=-2; korig=-2;lorig=-2
    orig=jorig+nxguard+nnx*(korig+nyguard)+(lorig+nzguard)*nnxy
    ngx=(ngridx-ncx)
    ngxy=(ngridx*ngridy-ncx*ncy)
    ncxy=ncx*ncy
    ! LOOP ON PARTICLES
    DO ip=1,np, LVEC
        !$OMP SIMD
        DO n=1,MIN(LVEC,np-ip+1)
            nn=ip+n-1
            ! --- computes position in  grid units at (n+1)
            x = (xp(nn)-xmin)*dxi
            y = (yp(nn)-ymin)*dyi
            z = (zp(nn)-zmin)*dzi

            ! Computes velocity
            usq = (uxp(nn)**2 + uyp(nn)**2+uzp(nn)**2)*clightsq
            gaminv = 1.0_num/sqrt(1.0_num + usq)
            vx = uxp(nn)*gaminv
            vy = uyp(nn)*gaminv
            vz = uzp(nn)*gaminv

            ! --- computes particles weights
            wq=q*w(nn)*invvol
            wqx(n)=wq*vx
            wqy(n)=wq*vy
            wqz(n)=wq*vz

            ! Gets position in grid units at (n+1/2) for computing rho(n+1/2)
            xmid=x-dts2dx*vx
            ymid=y-dts2dy*vy
            zmid=z-dts2dz*vz

            ! --- finds node of cell containing particles for current positions
            j=floor(xmid)
            k=floor(ymid)
            l=floor(zmid)
            j0=floor(xmid-0.5_num)
            k0=floor(ymid-0.5_num)
            l0=floor(zmid-0.5_num)
            ICELL(n,1)=1+(j0-jorig)+(k-korig)*ncx+(l-lorig)*ncxy
            ICELL(n,2)=1+(j-jorig)+(k0-korig)*ncx+(l-lorig)*ncxy
            ICELL(n,3)=1+(j-jorig)+(k-korig)*ncx+(l0-lorig)*ncxy

            ! --- computes set of coefficients for node centered quantities
            sx(n) = xmid-j
            sy(n) = ymid-k
            sz(n) = zmid-l

            ! --- computes set of coefficients for staggered quantities
            sx0(n) = xmid-j0-0.5_num
            sy0(n) = ymid-k0-0.5_num
            sz0(n) = zmid-l0-0.5_num
        END DO
        !$OMP END SIMD
        DO n=1,MIN(LVEC,np-ip+1)
            !$OMP SIMD
            DO nv=1,8
                wx=-mx(nv)+sx(n)
                wx0=-mx(nv)+sx0(n)
                wy=-my(nv)+sy(n)
                wy0=-my(nv)+sy0(n)
                wz=-mz(nv)+sz(n)
                wz0=-mz(nv)+sz0(n)
                wwx=wx0*wy*wz*wqx(n)*sgn(nv)
                wwy=wx*wy0*wz*wqy(n)*sgn(nv)
                wwz=wx*wy*wz0*wqz(n)*sgn(nv)
                ! --- add current contributions in the form rho(n+1/2)v(n+1/2)
                ! - JX
                jxcells(nv,ICELL(n,1))=jxcells(nv,ICELL(n,1))+wwx
                ! - JY
                jycells(nv,ICELL(n,2))=jycells(nv,ICELL(n,2))+wwy
                ! - JZ
                jzcells(nv,ICELL(n,3))=jzcells(nv,ICELL(n,3))+wwz
            END DO
            !$OMP END SIMD
        END DO
    END DO
    ! Reduction of jxcells,jycells,jzcells in jx,jy,jz
    DO iz=1, ncz
        DO iy=1,ncy
            !$OMP SIMD
            DO ix=1,ncx !! VECTOR (take ncx multiple of vector length)
                ic=ix+(iy-1)*ncx+(iz-1)*ncxy
                igrid=ic+(iy-1)*ngx+(iz-1)*ngxy
                ! jx
                jx(orig+igrid+moff(1))=jx(orig+igrid+moff(1))+jxcells(1,ic)
                jx(orig+igrid+moff(2))=jx(orig+igrid+moff(2))+jxcells(2,ic)
                jx(orig+igrid+moff(3))=jx(orig+igrid+moff(3))+jxcells(3,ic)
                jx(orig+igrid+moff(4))=jx(orig+igrid+moff(4))+jxcells(4,ic)
                jx(orig+igrid+moff(5))=jx(orig+igrid+moff(5))+jxcells(5,ic)
                jx(orig+igrid+moff(6))=jx(orig+igrid+moff(6))+jxcells(6,ic)
                jx(orig+igrid+moff(7))=jx(orig+igrid+moff(7))+jxcells(7,ic)
                jx(orig+igrid+moff(8))=jx(orig+igrid+moff(8))+jxcells(8,ic)
                ! jy
                jy(orig+igrid+moff(1))=jy(orig+igrid+moff(1))+jycells(1,ic)
                jy(orig+igrid+moff(2))=jy(orig+igrid+moff(2))+jycells(2,ic)
                jy(orig+igrid+moff(3))=jy(orig+igrid+moff(3))+jycells(3,ic)
                jy(orig+igrid+moff(4))=jy(orig+igrid+moff(4))+jycells(4,ic)
                jy(orig+igrid+moff(5))=jy(orig+igrid+moff(5))+jycells(5,ic)
                jy(orig+igrid+moff(6))=jy(orig+igrid+moff(6))+jycells(6,ic)
                jy(orig+igrid+moff(7))=jy(orig+igrid+moff(7))+jycells(7,ic)
                jy(orig+igrid+moff(8))=jy(orig+igrid+moff(8))+jycells(8,ic)
                ! jz
                jz(orig+igrid+moff(1))=jz(orig+igrid+moff(1))+jzcells(1,ic)
                jz(orig+igrid+moff(2))=jz(orig+igrid+moff(2))+jzcells(2,ic)
                jz(orig+igrid+moff(3))=jz(orig+igrid+moff(3))+jzcells(3,ic)
                jz(orig+igrid+moff(4))=jz(orig+igrid+moff(4))+jzcells(4,ic)
                jz(orig+igrid+moff(5))=jz(orig+igrid+moff(5))+jzcells(5,ic)
                jz(orig+igrid+moff(6))=jz(orig+igrid+moff(6))+jzcells(6,ic)
                jz(orig+igrid+moff(7))=jz(orig+igrid+moff(7))+jzcells(7,ic)
                jz(orig+igrid+moff(8))=jz(orig+igrid+moff(8))+jzcells(8,ic)
            END DO
            !$OMP END SIMD
        END DO
    END DO
    DEALLOCATE(jxcells,jycells,jzcells)
    RETURN
END SUBROUTINE depose_jxjyjz_vecHVv2_1_1_1\end{lstlisting}

\subsection{Order $2$ current deposition routine}
\begin{lstlisting}[frame=single,caption=New vector version of current deposition routine  for TSC particle shape factors,label=VecHVCurrentRoutineTSC, basicstyle=\ttfamily\footnotesize] 
SUBROUTINE depose_jxjyjz_vecHVv2_2_2_2(jx,jy,jz,np,xp,yp,zp,uxp,uyp,uzp,w,q, &
    xmin,ymin,zmin,dt,dx,dy,dz,nx,ny,nz,nxguard,nyguard,nzguard)
    USE constants
    IMPLICIT NONE
    INTEGER :: np,nx,ny,nz,nxguard,nyguard,nzguard
    REAL(num),INTENT(IN OUT) :: jx(1:(1+nx+2*nxguard)*(1+ny+2*nyguard)*(1+nz+2*nzguard))
    REAL(num),INTENT(IN OUT) :: jy(1:(1+nx+2*nxguard)*(1+ny+2*nyguard)*(1+nz+2*nzguard))
    REAL(num),INTENT(IN OUT) :: jz(1:(1+nx+2*nxguard)*(1+ny+2*nyguard)*(1+nz+2*nzguard))
    REAL(num), DIMENSION(:,:), ALLOCATABLE:: jxcells,jycells,jzcells
    REAL(num), DIMENSION(np) :: xp,yp,zp,uxp,uyp,uzp, w
    REAL(num) :: q,dt,dx,dy,dz,xmin,ymin,zmin
    REAL(num) :: dxi,dyi,dzi,xint,yint,zint, &
                   oxint,oyint,ozint,xintsq,yintsq,zintsq,oxintsq,oyintsq,ozintsq
    REAL(num) :: x,y,z,xmid,ymid,zmid,invvol, dts2dx, dts2dy, dts2dz
    REAL(num) ::   wqx,wqy,wqz,ww, wwx, wwy, wwz, gaminv, usq, clightsq
    REAL(num), PARAMETER :: onesixth=1.0_num/6.0_num,twothird=2.0_num/3.0_num
    INTEGER :: j,k,l,j0,k0,l0,ip, NCELLS, ic
    INTEGER :: nnx, nnxy, n,nn,nv
    INTEGER :: moff(1:8)
    INTEGER, PARAMETER :: LVEC=8
    INTEGER, DIMENSION(LVEC,3) :: ICELL, IG
    REAL(num) :: vx,vy,vz
    REAL(num) :: ww0x(LVEC,4),ww0y(LVEC,4),ww0z(LVEC,4), wwwx(LVEC,8), &
    wwwy(LVEC,8),wwwz(LVEC,8), wq
    REAL(num) :: sx0(LVEC),sx1(LVEC),sx2(LVEC)
    REAL(num) :: sx00(LVEC),sx01(LVEC),sx02(LVEC)
    REAL(num) :: sy0,sy1,sy2,sy00,sy01,sy02
    REAL(num) :: sz0,sz1,sz2,sz00,sz01,sz02, syz
    INTEGER :: igrid,orig, jorig, korig, lorig
    INTEGER :: ncx, ncy, ncxy, ncz,ix,iy,iz, ngridx, ngridy, ngx, ngxy

    dxi = 1.0_num/dx
    dyi = 1.0_num/dy
    dzi = 1.0_num/dz
    invvol = dxi*dyi*dzi
    dts2dx = 0.5_num*dt*dxi
    dts2dy = 0.5_num*dt*dyi
    dts2dz = 0.5_num*dt*dzi
    clightsq = 1.0_num/clight**2
    ww0x=0._num; ww0y=0._num; ww0z=0._num
    ngridx=nx+1+2*nxguard;ngridy=ny+1+2*nyguard
    ncx=nx+4;ncy=ny+4;ncz=nz+4
    NCELLS=ncx*ncy*ncz
    ALLOCATE(jxcells(8,NCELLS),jycells(8,NCELLS),jzcells(8,NCELLS))
    jxcells=0.0_num; jycells=0.0_num; jzcells=0.0_num
    nnx = nx + 1 + 2*nxguard
    nnxy = nnx*(ny+1+2*nyguard)
    moff = (/-nnx-nnxy,-nnxy,nnx-nnxy,-nnx,nnx,-nnx+nnxy,nnxy,nnx+nnxy/)
    jorig=-2; korig=-2;lorig=-2
    orig=jorig+nxguard+nnx*(korig+nyguard)+(lorig+nzguard)*nnxy
    ngx=(ngridx-ncx)
    ngxy=(ngridx*ngridy-ncx*ncy)
    ncxy=ncx*ncy
    ! LOOP ON PARTICLES
    DO ip=1,np, LVEC
        !$OMP SIMD
        DO n=1,MIN(LVEC,np-ip+1)
            nn=ip+n-1
            ! --- computes position in  grid units at (n+1)
            x = (xp(nn)-xmin)*dxi
            y = (yp(nn)-ymin)*dyi
            z = (zp(nn)-zmin)*dzi

            ! Computes velocity
            usq = (uxp(nn)**2 + uyp(nn)**2+uzp(nn)**2)*clightsq
            gaminv = 1.0_num/sqrt(1.0_num + usq)
            vx = uxp(nn)*gaminv
            vy = uyp(nn)*gaminv
            vz = uzp(nn)*gaminv

            ! --- computes particles weights
            wq=q*w(nn)*invvol
            wqx=wq*vx
            wqy=wq*vy
            wqz=wq*vz

            ! Gets position in grid units at (n+1/2) for computing rho(n+1/2)
            xmid=x-dts2dx*vx
            ymid=y-dts2dy*vy
            zmid=z-dts2dz*vz

            ! --- finds node of cell containing particles for current positions
            j=nint(xmid)
            k=nint(ymid)
            l=nint(zmid)
            j0=nint(xmid-0.5_num)
            k0=nint(ymid-0.5_num)
            l0=nint(zmid-0.5_num)
            ICELL(n,1)=1+(j0-jorig)+(k-korig)*ncx+(l-lorig)*ncxy
            ICELL(n,2)=1+(j-jorig)+(k0-korig)*ncx+(l-lorig)*ncxy
            ICELL(n,3)=1+(j-jorig)+(k-korig)*ncx+(l0-lorig)*ncxy
            IG(n,1)=ICELL(n,1)+(k-korig)*ngx+(l-lorig)*ngxy
            IG(n,2)=ICELL(n,2)+(k0-korig)*ngx+(l-lorig)*ngxy
            IG(n,3)=ICELL(n,3)+(k-korig)*ngx+(l0-lorig)*ngxy

            ! --- computes set of coefficients for node centered quantities
            xint = xmid-j
            yint = ymid-k
            zint = zmid-l
            xintsq= xint**2
            yintsq= yint**2
            zintsq= zint**2
            sx0(n)=0.5_num*(0.5_num-xint)**2
            sx1(n)=(0.75_num-xintsq)
            sx2(n)=0.5_num*(0.5_num+xint)**2
            sy0=0.5_num*(0.5_num-yint)**2
            sy1=(0.75_num-yintsq)
            sy2=0.5_num*(0.5_num+yint)**2
            sz0=0.5_num*(0.5_num-zint)**2
            sz1=(0.75_num-zintsq)
            sz2=0.5_num*(0.5_num+zint)**2

            ! --- computes set of coefficients for staggered quantities
            xint = xmid-j0-0.5_num
            yint = ymid-k0-0.5_num
            zint = zmid-l0-0.5_num
            xintsq= xint**2
            yintsq= yint**2
            zintsq= zint**2
            sx00(n)=0.5_num*(0.5_num-xint)**2
            sx01(n)=(0.75_num-xintsq)
            sx02(n)=0.5_num*(0.5_num+xint)**2
            sy00=0.5_num*(0.5_num-yint)**2
            sy01=(0.75_num-yintsq)
            sy02=0.5_num*(0.5_num+yint)**2
            sz00=0.5_num*(0.5_num-zint)**2
            sz01=(0.75_num-zintsq)
            sz02=0.5_num*(0.5_num+zint)**2

            ! -- Weights for planes of 8  vertices
            ! Weights - X
            wwwx(n,1) = sy0*sz0*wqx
            wwwx(n,2) = sy1*sz0*wqx
            wwwx(n,3) = sy2*sz0*wqx
            wwwx(n,4) = sy0*sz1*wqx
            wwwx(n,5) = sy2*sz1*wqx
            wwwx(n,6) = sy0*sz2*wqx
            wwwx(n,7) = sy1*sz2*wqx
            wwwx(n,8) = sy2*sz2*wqx

            ! Weights - Y
            wwwy(n,1) = sy00*sz0*wqy
            wwwy(n,2) = sy01*sz0*wqy
            wwwy(n,3) = sy02*sz0*wqy
            wwwy(n,4) = sy00*sz1*wqy
            wwwy(n,5) = sy02*sz1*wqy
            wwwy(n,6) = sy00*sz2*wqy
            wwwy(n,7) = sy01*sz2*wqy
            wwwy(n,8) = sy02*sz2*wqy

            ! Weights - Z
            wwwz(n,1) = sy0*sz00*wqz
            wwwz(n,2) = sy1*sz00*wqz
            wwwz(n,3) = sy2*sz00*wqz
            wwwz(n,4) = sy0*sz01*wqz
            wwwz(n,5) = sy2*sz01*wqz
            wwwz(n,6) = sy0*sz02*wqz
            wwwz(n,7) = sy1*sz02*wqz
            wwwz(n,8) = sy2*sz02*wqz

            ! -- 3 remaining central points
            syz=sz1*sy1*wqx
            ww0x(n,1)=syz*sx00(n)
            ww0x(n,2)=syz*sx01(n)
            ww0x(n,3)=syz*sx02(n)
            syz=sz1*sy01*wqy
            ww0y(n,1)=syz*sx0(n)
            ww0y(n,2)=syz*sx1(n)
            ww0y(n,3)=syz*sx2(n)
            syz=sz01*sy1*wqz
            ww0z(n,1)=syz*sx0(n)
            ww0z(n,2)=syz*sx1(n)
            ww0z(n,3)=syz*sx2(n)
        END DO
        !$OMP END SIMD
        DO n=1,MIN(LVEC,np-ip+1)
            !$OMP SIMD
            DO nv=1,8
                ! --- add current contributions in the form rho(n+1/2)v(n+1/2)
                ! - JX
                wwx=wwwx(n,nv)
                ! Loop on (i=-1,j,k)
                jxcells(nv,ICELL(n,1)-1) = jxcells(nv,ICELL(n,1)-1) +wwx*sx00(n)
                ! Loop on (i=0,j,k)
                jxcells(nv,ICELL(n,1))   = jxcells(nv,ICELL(n,1))   +wwx*sx01(n)
                !Loop on (i=1,j,k)
                jxcells(nv,ICELL(n,1)+1) = jxcells(nv,ICELL(n,1)+1) +wwx*sx02(n)
                ! - JY
                wwy=wwwy(n,nv)
                ! Loop on (i=-1,j,k)
                jycells(nv,ICELL(n,2)-1) = jycells(nv,ICELL(n,2)-1) +wwy*sx0(n)
                ! Loop on (i=0,j,k)
                jycells(nv,ICELL(n,2))   = jycells(nv,ICELL(n,2))   +wwy*sx1(n)
                !Loop on (i=1,j,k)
                jycells(nv,ICELL(n,2)+1) = jycells(nv,ICELL(n,2)+1) +wwy*sx2(n)
                ! - JZ
                wwz=wwwz(n,nv)
                ! Loop on (i=-1,j,k)
                jzcells(nv,ICELL(n,3)-1) = jzcells(nv,ICELL(n,3)-1) +wwz*sx0(n)
                ! Loop on (i=0,j,k)
                jzcells(nv,ICELL(n,3))   = jzcells(nv,ICELL(n,3))   +wwz*sx1(n)
                !Loop on (i=1,j,k)
                jzcells(nv,ICELL(n,3)+1) = jzcells(nv,ICELL(n,3)+1) +wwz*sx2(n)
            END DO
            !$OMP END SIMD
            !$OMP SIMD
            DO nv=1,4
                jx(orig+IG(n,1)+nv-2)=jx(orig+IG(n,1)+nv-2)+ww0x(n,nv)
                jy(orig+IG(n,2)+nv-2)=jy(orig+IG(n,2)+nv-2)+ww0y(n,nv)
                jz(orig+IG(n,3)+nv-2)=jz(orig+IG(n,3)+nv-2)+ww0z(n,nv)
            END DO
            !$OMP END SIMD
        END DO
    END DO
    ! Reduction of jxcells,jycells,jzcells in jx,jy,jz
    DO iz=1, ncz
        DO iy=1,ncy
            !$OMP SIMD
            DO ix=1,ncx !! VECTOR (take ncx multiple of vector length)
                ic=ix+(iy-1)*ncx+(iz-1)*ncxy
                igrid=ic+(iy-1)*ngx+(iz-1)*ngxy
                ! jx
                jx(orig+igrid+moff(1))=jx(orig+igrid+moff(1))+jxcells(1,ic)
                jx(orig+igrid+moff(2))=jx(orig+igrid+moff(2))+jxcells(2,ic)
                jx(orig+igrid+moff(3))=jx(orig+igrid+moff(3))+jxcells(3,ic)
                jx(orig+igrid+moff(4))=jx(orig+igrid+moff(4))+jxcells(4,ic)
                jx(orig+igrid+moff(5))=jx(orig+igrid+moff(5))+jxcells(5,ic)
                jx(orig+igrid+moff(6))=jx(orig+igrid+moff(6))+jxcells(6,ic)
                jx(orig+igrid+moff(7))=jx(orig+igrid+moff(7))+jxcells(7,ic)
                jx(orig+igrid+moff(8))=jx(orig+igrid+moff(8))+jxcells(8,ic)
                ! jy
                jy(orig+igrid+moff(1))=jy(orig+igrid+moff(1))+jycells(1,ic)
                jy(orig+igrid+moff(2))=jy(orig+igrid+moff(2))+jycells(2,ic)
                jy(orig+igrid+moff(3))=jy(orig+igrid+moff(3))+jycells(3,ic)
                jy(orig+igrid+moff(4))=jy(orig+igrid+moff(4))+jycells(4,ic)
                jy(orig+igrid+moff(5))=jy(orig+igrid+moff(5))+jycells(5,ic)
                jy(orig+igrid+moff(6))=jy(orig+igrid+moff(6))+jycells(6,ic)
                jy(orig+igrid+moff(7))=jy(orig+igrid+moff(7))+jycells(7,ic)
                jy(orig+igrid+moff(8))=jy(orig+igrid+moff(8))+jycells(8,ic)
                ! jz
                jz(orig+igrid+moff(1))=jz(orig+igrid+moff(1))+jzcells(1,ic)
                jz(orig+igrid+moff(2))=jz(orig+igrid+moff(2))+jzcells(2,ic)
                jz(orig+igrid+moff(3))=jz(orig+igrid+moff(3))+jzcells(3,ic)
                jz(orig+igrid+moff(4))=jz(orig+igrid+moff(4))+jzcells(4,ic)
                jz(orig+igrid+moff(5))=jz(orig+igrid+moff(5))+jzcells(5,ic)
                jz(orig+igrid+moff(6))=jz(orig+igrid+moff(6))+jzcells(6,ic)
                jz(orig+igrid+moff(7))=jz(orig+igrid+moff(7))+jzcells(7,ic)
                jz(orig+igrid+moff(8))=jz(orig+igrid+moff(8))+jzcells(8,ic)
            END DO
            !$OMP END SIMD
        END DO
    END DO
    DEALLOCATE(jxcells,jycells,jzcells)
    RETURN
END SUBROUTINE depose_jxjyjz_vecHVv2_2_2_2
\end{lstlisting}

\subsection{Order $3$ current deposition routine}
\begin{lstlisting}[frame=single,caption=New vector version of current deposition routine  for QSP particle shape factors,label=VecHVCurrentRoutineQSP, basicstyle=\ttfamily\footnotesize] 
!!! Use with nox=4
SUBROUTINE depose_jxjyjz_vecHVv2_3_3_3(jx,jy,jz,np,xp,yp,zp,uxp,uyp,uzp,w,q, &
    xmin,ymin,zmin,dt,dx,dy,dz,nx,ny,nz,nxguard,nyguard,nzguard)
    USE constants
    IMPLICIT NONE
    INTEGER :: np,nx,ny,nz,nxguard,nyguard,nzguard
    REAL(num),INTENT(IN OUT) :: jx(1:(1+nx+2*nxguard)*(1+ny+2*nyguard)*(1+nz+2*nzguard))
    REAL(num),INTENT(IN OUT) :: jy(1:(1+nx+2*nxguard)*(1+ny+2*nyguard)*(1+nz+2*nzguard))
    REAL(num),INTENT(IN OUT) :: jz(1:(1+nx+2*nxguard)*(1+ny+2*nyguard)*(1+nz+2*nzguard))
    REAL(num), DIMENSION(:,:), ALLOCATABLE:: jxcells,jycells,jzcells
    REAL(num), DIMENSION(np) :: xp,yp,zp,uxp,uyp,uzp, w
    REAL(num) :: q,dt,dx,dy,dz,xmin,ymin,zmin
    REAL(num) :: dxi,dyi,dzi,xint,yint,zint, &
                   oxint,oyint,ozint,xintsq,yintsq,zintsq, oxintsq,oyintsq, ozintsq
    REAL(num) :: x,y,z,xmid,ymid,zmid,invvol, dts2dx, dts2dy, dts2dz
    REAL(num) ::   ww, wwx, wwy, wwz, gaminv, usq, clightsq
    REAL(num), PARAMETER :: onesixth=1.0_num/6.0_num,twothird=2.0_num/3.0_num
    INTEGER :: j,k,l,j0,k0,l0,ip, NCELLS, ic, ix, iy, iz
    INTEGER :: nnx, nnxy,ngridx, ngridy, n,nn,nv
    INTEGER :: moff(1:8)
    INTEGER, PARAMETER :: LVEC=8
    INTEGER, DIMENSION(LVEC,3) :: ICELL
    REAL(num), DIMENSION(LVEC) :: vx,vy,vz
    REAL(num) ::  wwwx(LVEC,16), wwwy(LVEC,16),wwwz(LVEC,16), wq
    REAL(num) :: sx1(LVEC),sx2(LVEC),sx3(LVEC),sx4(LVEC)
    REAL(num) :: sx01(LVEC),sx02(LVEC),sx03(LVEC),sx04(LVEC)
    REAL(num) :: sy1,sy2,sy3,sy4,sz1,sz2,sz3,sz4
    REAL(num) :: sy01,sy02,sy03,sy04,sz01,sz02,sz03,sz04
    REAL(num), DIMENSION(4) :: szz, zdec, h1, h11, h12, sgn
    REAL(num):: wwwx1(LVEC,8),wwwx2(LVEC,8),wwwy1(LVEC,8), &
    wwwy2(LVEC,8),wwwz1(LVEC,8),wwwz2(LVEC,8)
    REAL(num):: wx1,wx2,wy1,wy2,wz1,wz2
    INTEGER :: orig, ncxy, ncx, ncy, ncz, ngx, ngxy, igrid, jorig, korig, lorig

    dxi = 1.0_num/dx
    dyi = 1.0_num/dy
    dzi = 1.0_num/dz
    invvol = dxi*dyi*dzi
    dts2dx = 0.5_num*dt*dxi
    dts2dy = 0.5_num*dt*dyi
    dts2dz = 0.5_num*dt*dzi
    clightsq = 1.0_num/clight**2
    ngridx=nx+1+2*nxguard;ngridy=ny+1+2*nyguard
    ncx=nx+5; ncy=ny+4; ncz=nz+3
    NCELLS=ncx*ncy*ncz
    ALLOCATE(jxcells(8,NCELLS),jycells(8,NCELLS),jzcells(8,NCELLS))
    jxcells=0.0_num; jycells=0.0_num; jzcells=0.0_num;
    nnx = ngridx
    nnxy = ngridx*ngridy
    moff = (/-nnxy,0,nnxy,2*nnxy,nnx-nnxy,nnx,nnx+nnxy,nnx+2*nnxy/)
    jorig=-2; korig=-2;lorig=-2
    orig=jorig+nxguard+nnx*(korig+nyguard)+(lorig+nzguard)*nnxy
    ngx=(ngridx-ncx)
    ngxy=(ngridx*ngridy-ncx*ncy)
    ncxy=ncx*ncy

    h1=(/1_num,0_num,1_num,0_num/); sgn=(/1_num,-1_num,1_num,-1_num/)
    h11=(/0_num,1_num,1_num,0_num/); h12=(/1_num,0_num,0_num,1_num/)
    ! LOOP ON PARTICLES
    DO ip=1,np, LVEC
        !$OMP SIMD
        DO n=1,MIN(LVEC,np-ip+1)
            nn=ip+n-1
            ! --- computes position in  grid units at (n+1)
            x = (xp(nn)-xmin)*dxi
            y = (yp(nn)-ymin)*dyi
            z = (zp(nn)-zmin)*dzi

            ! Computes velocity
            usq = (uxp(nn)**2 + uyp(nn)**2+uzp(nn)**2)*clightsq
            gaminv = 1.0_num/sqrt(1.0_num + usq)
            vx(n) = uxp(nn)*gaminv
            vy(n) = uyp(nn)*gaminv
            vz(n) = uzp(nn)*gaminv

            ! --- computes particles weights
            wq=q*w(nn)*invvol

            ! Gets position in grid units at (n+1/2) for computing rho(n+1/2)
            xmid=x-dts2dx*vx(n)
            ymid=y-dts2dy*vy(n)
            zmid=z-dts2dz*vz(n)

            ! --- finds node of cell containing particles for current positions
            j=floor(xmid)
            k=floor(ymid)
            l=floor(zmid)
            j0=floor(xmid-0.5_num)
            k0=floor(ymid-0.5_num)
            l0=floor(zmid-0.5_num)
            ICELL(n,1)=1+(j0-jorig)+(k-korig)*ncx+(l-lorig)*ncxy
            ICELL(n,2)=1+(j-jorig)+(k0-korig)*ncx+(l-lorig)*ncxy
            ICELL(n,3)=1+(j-jorig)+(k-korig)*ncx+(l0-lorig)*ncxy

            ! --- computes set of coefficients for node centered quantities
            xint    = xmid-j
            yint    = ymid-k
            zint    = zmid-l
            oxint   = 1.0_num-xint
            xintsq  = xint*xint
            oxintsq = oxint*oxint
            sx1(n)  = onesixth*oxintsq*oxint
            sx2(n)  = twothird-xintsq*(1.0_num-xint*0.5_num)
            sx3(n)  = twothird-oxintsq*(1.0_num-oxint*0.5_num)
            sx4(n)  = onesixth*xintsq*xint
            oyint   = 1.0_num-yint
            yintsq  = yint*yint
            oyintsq = oyint*oyint
            sy1  = onesixth*oyintsq*oyint
            sy2  = (twothird-yintsq*(1.0_num-yint*0.5_num))
            sy3  = (twothird-oyintsq*(1.0_num-oyint*0.5_num))
            sy4  = onesixth*yintsq*yint
            ozint = 1.0_num-zint
            zintsq = zint*zint
            ozintsq = ozint*ozint
            sz1 = onesixth*ozintsq*ozint*wq
            sz2 = (twothird-zintsq*(1.0_num-zint*0.5_num))*wq
            sz3 = (twothird-ozintsq*(1.0_num-ozint*0.5_num))*wq
            sz4 = onesixth*zintsq*zint*wq

            ! --- computes set of coefficients for staggered quantities
            xint     = xmid-j0-0.5_num
            yint     = ymid-k0-0.5_num
            zint     = zmid-l0-0.5_num
            oxint    = 1.0_num-xint
            xintsq   = xint*xint
            oxintsq  = oxint*oxint
            sx01(n)  = onesixth*oxintsq*oxint
            sx02(n)  = twothird-xintsq*(1.0_num-xint*0.5_num)
            sx03(n)  = twothird-oxintsq*(1.0_num-oxint*0.5_num)
            sx04(n)  = onesixth*xintsq*xint
            oyint    = 1.0_num-yint
            yintsq   = yint*yint
            oyintsq  = oyint*oyint
            sy01  = onesixth*oyintsq*oyint
            sy02  = (twothird-yintsq*(1.0_num-yint*0.5_num))
            sy03  = (twothird-oyintsq*(1.0_num-oyint*0.5_num))
            sy04  = onesixth*yintsq*yint
            ozint = 1.0_num-zint
            zintsq = zint*zint
            ozintsq = ozint*ozint
            sz01 = onesixth*ozintsq*ozint*wq
            sz02 = (twothird-zintsq*(1.0_num-zint*0.5_num))*wq
            sz03 = (twothird-ozintsq*(1.0_num-ozint*0.5_num))*wq
            sz04 = onesixth*zintsq*zint*wq
            ! --- computes weights
            ! - X
            wwwx1(n,1)=sz1*sy1
            wwwx1(n,2)=sz2*sy1
            wwwx1(n,3)=sz3*sy1
            wwwx1(n,4)=sz4*sy1
            wwwx1(n,5)=sz1*sy2
            wwwx1(n,6)=sz2*sy2
            wwwx1(n,7)=sz3*sy2
            wwwx1(n,8)=sz4*sy2
            wwwx2(n,1)=sz1*sy3
            wwwx2(n,2)=sz2*sy3
            wwwx2(n,3)=sz3*sy3
            wwwx2(n,4)=sz4*sy3
            wwwx2(n,5)=sz1*sy4
            wwwx2(n,6)=sz2*sy4
            wwwx2(n,7)=sz3*sy4
            wwwx2(n,8)=sz4*sy4
            ! - Y
            wwwy1(n,1)=sz1*sy01
            wwwy1(n,2)=sz2*sy01
            wwwy1(n,3)=sz3*sy01
            wwwy1(n,4)=sz4*sy01
            wwwy1(n,5)=sz1*sy02
            wwwy1(n,6)=sz2*sy02
            wwwy1(n,7)=sz3*sy02
            wwwy1(n,8)=sz4*sy02
            wwwy2(n,1)=sz1*sy03
            wwwy2(n,2)=sz2*sy03
            wwwy2(n,3)=sz3*sy03
            wwwy2(n,4)=sz4*sy03
            wwwy2(n,5)=sz1*sy04
            wwwy2(n,6)=sz2*sy04
            wwwy2(n,7)=sz3*sy04
            wwwy2(n,8)=sz4*sy04
            ! - Y
            wwwy1(n,1)=sz1*sy01
            wwwy1(n,2)=sz2*sy01
            wwwy1(n,3)=sz3*sy01
            wwwy1(n,4)=sz4*sy01
            wwwy1(n,5)=sz1*sy02
            wwwy1(n,6)=sz2*sy02
            wwwy1(n,7)=sz3*sy02
            wwwy1(n,8)=sz4*sy02
            wwwy2(n,1)=sz1*sy03
            wwwy2(n,2)=sz2*sy03
            wwwy2(n,3)=sz3*sy03
            wwwy2(n,4)=sz4*sy03
            wwwy2(n,5)=sz1*sy04
            wwwy2(n,6)=sz2*sy04
            wwwy2(n,7)=sz3*sy04
            wwwy2(n,8)=sz4*sy04
            ! - Y
            wwwz1(n,1)=sz01*sy1
            wwwz1(n,2)=sz02*sy1
            wwwz1(n,3)=sz03*sy1
            wwwz1(n,4)=sz04*sy1
            wwwz1(n,5)=sz01*sy2
            wwwz1(n,6)=sz02*sy2
            wwwz1(n,7)=sz03*sy2
            wwwz1(n,8)=sz04*sy2
            wwwz2(n,1)=sz01*sy3
            wwwz2(n,2)=sz02*sy3
            wwwz2(n,3)=sz03*sy3
            wwwz2(n,4)=sz04*sy3
            wwwz2(n,5)=sz01*sy4
            wwwz2(n,6)=sz02*sy4
            wwwz2(n,7)=sz03*sy4
            wwwz2(n,8)=sz04*sy4
        END DO
        !$OMP END SIMD

        ! Add weights to nearest vertices
        DO n=1,MIN(LVEC,np-ip+1)
            !$OMP SIMD
            DO nv=1,8
                ! --- JX
                wx1=wwwx1(n,nv); wx2=wwwx2(n,nv)
                ! Loop on (i=-1,j,k)
                jxcells(nv,ICELL(n,1)-ncx-1) = jxcells(nv,ICELL(n,1)-ncx-1) + &
                 wx1*sx01(n)*vx(n)
                ! Loop on (i=0,j,k)
                jxcells(nv,ICELL(n,1)-ncx)   = jxcells(nv,ICELL(n,1)-ncx)   + &
                 wx1*sx02(n)*vx(n)
                !Loop on (i=1,j,k)
                jxcells(nv,ICELL(n,1)-ncx+1) = jxcells(nv,ICELL(n,1)-ncx+1) + &
                wx1*sx03(n)*vx(n)
                !Loop on (i=1,j,k)
                jxcells(nv,ICELL(n,1)-ncx+2) = jxcells(nv,ICELL(n,1)-ncx+2) + &
                wx1*sx04(n)*vx(n)
                ! Loop on (i=-1,j,k)
                jxcells(nv,ICELL(n,1)+ncx-1) = jxcells(nv,ICELL(n,1)+ncx-1) + &
                wx2*sx01(n)*vx(n)
                ! Loop on (i=0,j,k)
                jxcells(nv,ICELL(n,1)+ncx)   = jxcells(nv,ICELL(n,1)+ncx)   + &
                wx2*sx02(n)*vx(n)
                !Loop on (i=1,j,k)
                jxcells(nv,ICELL(n,1)+ncx+1) = jxcells(nv,ICELL(n,1)+ncx+1) + &
                wx2*sx03(n)*vx(n)
                !Loop on (i=1,j,k)
                jxcells(nv,ICELL(n,1)+ncx+2) = jxcells(nv,ICELL(n,1)+ncx+2) + &
                wx2*sx04(n)*vx(n)

                ! --- JY
                wy1=wwwy1(n,nv); wy2=wwwy2(n,nv)
                ! Loop on (i=-1,j,k)
                jycells(nv,ICELL(n,2)-ncx-1) = jycells(nv,ICELL(n,2)-ncx-1) + &
                wy1*sx1(n)*vy(n)
                ! Loop on (i=0,j,k)
                jycells(nv,ICELL(n,2)-ncx)   = jycells(nv,ICELL(n,2)-ncx)   + &
                wy1*sx2(n)*vy(n)
                !Loop on (i=1,j,k)
                jycells(nv,ICELL(n,2)-ncx+1) = jycells(nv,ICELL(n,2)-ncx+1) + &
                wy1*sx3(n)*vy(n)
                !Loop on (i=1,j,k)
                jycells(nv,ICELL(n,2)-ncx+2) = jycells(nv,ICELL(n,2)-ncx+2) + &
                wy1*sx4(n)*vy(n)
                ! Loop on (i=-1,j,k)
                jycells(nv,ICELL(n,2)+ncx-1) = jycells(nv,ICELL(n,2)+ncx-1) + &
                wy2*sx1(n)*vy(n)
                ! Loop on (i=0,j,k)
                jycells(nv,ICELL(n,2)+ncx)   = jycells(nv,ICELL(n,2)+ncx)   + &
                wy2*sx2(n)*vy(n)
                !Loop on (i=1,j,k)
                jycells(nv,ICELL(n,2)+ncx+1) = jycells(nv,ICELL(n,2)+ncx+1) + &
                wy2*sx3(n)*vy(n)
                !Loop on (i=1,j,k)
                jycells(nv,ICELL(n,2)+ncx+2) = jycells(nv,ICELL(n,2)+ncx+2) + &
                wy2*sx4(n)*vy(n)

                ! --- JZ
                wz1=wwwz1(n,nv); wz2=wwwz2(n,nv)
                ! Loop on (i=-1,j,k)
                jzcells(nv,ICELL(n,3)-ncx-1) = jzcells(nv,ICELL(n,3)-ncx-1) + &
                wz1*sx1(n)*vz(n)
                ! Loop on (i=0,j,k)
                jzcells(nv,ICELL(n,3)-ncx)   = jzcells(nv,ICELL(n,3)-ncx)   + &
                wz1*sx2(n)*vz(n)
                !Loop on (i=1,j,k)
                jzcells(nv,ICELL(n,3)-ncx+1) = jzcells(nv,ICELL(n,3)-ncx+1) + &
                wz1*sx3(n)*vz(n)
                !Loop on (i=1,j,k)
                jzcells(nv,ICELL(n,3)-ncx+2) = jzcells(nv,ICELL(n,3)-ncx+2) + &
                wz1*sx4(n)*vz(n)
                ! Loop on (i=-1,j,k)
                jzcells(nv,ICELL(n,3)+ncx-1) = jzcells(nv,ICELL(n,3)+ncx-1) + &
                wz2*sx1(n)*vz(n)
                ! Loop on (i=0,j,k)
                jzcells(nv,ICELL(n,3)+ncx)   = jzcells(nv,ICELL(n,3)+ncx)   + &
                wz2*sx2(n)*vz(n)
                !Loop on (i=1,j,k)
                jzcells(nv,ICELL(n,3)+ncx+1) = jzcells(nv,ICELL(n,3)+ncx+1) + &
                wz2*sx3(n)*vz(n)
                !Loop on (i=1,j,k)
                jzcells(nv,ICELL(n,3)+ncx+2) = jzcells(nv,ICELL(n,3)+ncx+2) + &
                wz2*sx4(n)*vz(n)
            END DO
            !$OMP END SIMD
        END DO
    END DO
    ! Reduction of jxcells,jycells,jzcells in jx,jy,jz
    DO iz=1, ncz
        DO iy=1,ncy
            !$OMP SIMD
            DO ix=1,ncx !! VECTOR (take ncx multiple of vector length)
                ic=ix+(iy-1)*ncx+(iz-1)*ncxy
                igrid=ic+(iy-1)*ngx+(iz-1)*ngxy
                ! jx
                jx(orig+igrid+moff(1))=jx(orig+igrid+moff(1))+jxcells(1,ic)
                jx(orig+igrid+moff(2))=jx(orig+igrid+moff(2))+jxcells(2,ic)
                jx(orig+igrid+moff(3))=jx(orig+igrid+moff(3))+jxcells(3,ic)
                jx(orig+igrid+moff(4))=jx(orig+igrid+moff(4))+jxcells(4,ic)
                jx(orig+igrid+moff(5))=jx(orig+igrid+moff(5))+jxcells(5,ic)
                jx(orig+igrid+moff(6))=jx(orig+igrid+moff(6))+jxcells(6,ic)
                jx(orig+igrid+moff(7))=jx(orig+igrid+moff(7))+jxcells(7,ic)
                jx(orig+igrid+moff(8))=jx(orig+igrid+moff(8))+jxcells(8,ic)
                ! jy
                jy(orig+igrid+moff(1))=jy(orig+igrid+moff(1))+jycells(1,ic)
                jy(orig+igrid+moff(2))=jy(orig+igrid+moff(2))+jycells(2,ic)
                jy(orig+igrid+moff(3))=jy(orig+igrid+moff(3))+jycells(3,ic)
                jy(orig+igrid+moff(4))=jy(orig+igrid+moff(4))+jycells(4,ic)
                jy(orig+igrid+moff(5))=jy(orig+igrid+moff(5))+jycells(5,ic)
                jy(orig+igrid+moff(6))=jy(orig+igrid+moff(6))+jycells(6,ic)
                jy(orig+igrid+moff(7))=jy(orig+igrid+moff(7))+jycells(7,ic)
                jy(orig+igrid+moff(8))=jy(orig+igrid+moff(8))+jycells(8,ic)
                ! jz
                jz(orig+igrid+moff(1))=jz(orig+igrid+moff(1))+jzcells(1,ic)
                jz(orig+igrid+moff(2))=jz(orig+igrid+moff(2))+jzcells(2,ic)
                jz(orig+igrid+moff(3))=jz(orig+igrid+moff(3))+jzcells(3,ic)
                jz(orig+igrid+moff(4))=jz(orig+igrid+moff(4))+jzcells(4,ic)
                jz(orig+igrid+moff(5))=jz(orig+igrid+moff(5))+jzcells(5,ic)
                jz(orig+igrid+moff(6))=jz(orig+igrid+moff(6))+jzcells(6,ic)
                jz(orig+igrid+moff(7))=jz(orig+igrid+moff(7))+jzcells(7,ic)
                jz(orig+igrid+moff(8))=jz(orig+igrid+moff(8))+jzcells(8,ic)
            END DO
            !$OMP END SIMD
        END DO
    END DO
    DEALLOCATE(jxcells,jycells,jzcells)
    RETURN
END SUBROUTINE depose_jxjyjz_vecHVv2_3_3_3
\end{lstlisting}
\end{document}